\documentclass[12pt]{cernart}
\usepackage{epsfig}
\usepackage{graphicx} 
\usepackage{amsmath,amssymb}
\newcommand{\defmath}[2] {\def#1{\ifmmode{#2}\else\mbox{${#2}$}\fi}}
\newcommand{\defdecay}[2] {\def#1{\ifmmode{#2}\else\mbox{${#2}$}\fi}}

\newcommand{\defunit}[2] {\def#1{\ifmmode\mathrm{#2}\else\mbox{$\mathrm{#2}$}\fi
}} % easy to define math sequences for units

\defunit{\mum}{\mu m}
\defunit{\mus}{\mu s}
\defunit{\taus}{\tau_S}
\defunit{\degrees}{^{\circ}}
\defunit{\xo}{X_0}

\defdecay{\k} {K}
\defdecay{\kz} {K^0}
\defdecay{\kbar} {\overline{K^{0}}}
\defdecay{\cascadebar} {\overline{\Xi^0}}
\defdecay{\lambdabar} {\overline{\Lambda}}
\defdecay{\kone} {K_1}
\defdecay{\ktwo} {K_2}
\defdecay{\ks} {K_S}
\defdecay{\kl} {K_L}
\defdecay{\ksl} {K_{S,L}}
\defdecay{\KL} {\kl}
\defdecay{\KS} {\ks}
\defdecay{\sbar} {\overline{\mathrm{s}}}
\defdecay{\dbar} {\overline{\mathrm{d}}}

\defdecay{\pipi} {\pi^{+}\pi^{-}}
\defdecay{\twopi} {2\pi}
\defdecay{\kpipi} {\k \rightarrow \pipi}
\defdecay{\ktwopi} {\ks \rightarrow \twopi}
\defdecay{\kspipi} {\ks \rightarrow \pipi}
\defdecay{\kstwopi} {\kl \rightarrow \twopi}
\defdecay{\klpipi} {\kl \rightarrow \pipi}
\defdecay{\kltwopi} {\k \rightarrow \twopi}
\defdecay{\pipipi} {\pi\pi\pi}
\defdecay{\threepi} {3\pi}

\defdecay{\pz}{\pi^0}
\defdecay{\pio} {\pi^0}
\defdecay{\twopio} {2 \pi^0}
\defdecay{\piopio} {\pi^0 \pi^0}
\defdecay{\pipin} {\piopio}
\defdecay{\ktwopio} {\k \rightarrow \twopio}
\defdecay{\kpiopio} {\k \rightarrow \piopio}
\defdecay{\kltwopio} {\kl \rightarrow \twopio}
\defdecay{\klpiopio} {\kl \rightarrow \piopio}
\defdecay{\kstwopio} {\ks \rightarrow \twopio}
\defdecay{\kspiopio} {\ks \rightarrow \piopio}
\defdecay{\klthreepich} {\kl \rightarrow \pio\pi^{+}\pi^{-}}

\defdecay{\kspiee} {\ks \rightarrow \pio e^+ e^-}
\defdecay{\klpiee} {\kl \rightarrow \pio e^+ e^-}
\defdecay{\kspipiD} {\ks \rightarrow \pio \pioD }
\defdecay{\kspipid} {\ks \rightarrow \pio \piod }
\defdecay{\kspiDpiD} {\ks \rightarrow \pioD \pioD }
\defdecay{\kspidpid} {\ks \rightarrow \piod \piod }
\defdecay{\kspipidd} {\ks \rightarrow \pio \piodd}
\defdecay{\kpiee} {K \rightarrow \pi e e}
\defdecay{\piee} {\pio e^+ e^-}
\defdecay{\pipid} {\pio \piod}
\defdecay{\pidpid} {\piod \piod}
\defdecay{\pipidd} {\pio \piodd}
\defdecay{\p0pipi} {\pio \pi^{+} \pi^{-} }
\defdecay{\kleegg} {\kl\rightarrow e e\gamma\gamma}
\defdecay{\klpigg} {\kl\rightarrow \pio \gamma \gamma}
\defdecay{\kseegg} {\ks\rightarrow e e\gamma\gamma}
\defdecay{\eegg} {e e\gamma\gamma}
\defdecay{\kleeg} {\kl\rightarrow e e\gamma}
\defdecay{\klp0pipi} {\kl\rightarrow \pio \pi^+ \pi^- }
\defdecay{\BRkspiee} {B ( \ks \rightarrow \pio e^+ e^- ) }
\defdecay{\BRklpiee} {B ( \kl \rightarrow \pio e^+ e^- ) }
\defdecay{\BRkspipid} {B ( \kspipid ) }
\defdecay{\BRkspidpid} {B ( \kspidpid ) }
\defdecay{\BRklp0pipi} {B ( \klp0pipi) }
\defdecay{\BRp0pipi} {B ( \p0pipi) }
\defdecay{\Xilampi} {\Xi^0 \rightarrow \Lambda \pio}
\defdecay{\lamppi} {\Lambda \rightarrow p \pi^{-}}
\defdecay{\Xilamppipio} {\Xi^{0}\rightarrow\Lambda (p \pi^{-})\pio }
\defdecay{\lampev} {\Lambda \rightarrow p e^{-} \nu}
\defdecay{\Xilampevpio} {\Xi^{0}\rightarrow\Lambda (p e^{-} \nu)\pio }
\defdecay{\Xisigmaenv} {\Xi^0 \rightarrow \Sigma^{+} e^{-} \nu}
\defdecay{\sigppio} {\Sigma^{+} \rightarrow  p \pio}
\defdecay{\Xisigmappioenv} {\Xi^0 \rightarrow \Sigma^{+}(p \pio) e^{-} \nu}
\defdecay{\AXilampi} {\overline{\Xi^0}\rightarrow \overline{\Lambda} \pio  }
\defdecay{\Alamppi} {\overline{\Lambda} \rightarrow \overline{p} \pi^{+}}
\defdecay{\Alampev} {\overline{\Lambda} \rightarrow \overline{p} e^{+} \, \overline{\nu}}
\defdecay{\AXisigmaenv} {\overline{\Xi^0} \rightarrow \overline{\Sigma^{+}} e^{+} \nu}
\defdecay{\Asigppio} {\overline{\Sigma^{+}} \rightarrow  \overline{p} \pio}
\defdecay{\klkef} {\kl\rightarrow \pio \pi^{\pm} e^{\mp} \nu }
\defdecay{\kef} {\pio \pi^{\pm} e^{\mp} \nu }
\defdecay{\klketh} {\kl\rightarrow \pi^{\pm} e^{\mp} \nu }
\defdecay{\keth} {\pi^{\pm} e^{\mp} \nu }

\defdecay{\dalconv} {{\bf\boldmath \pio}(\gamma\gamma){\bf\boldmath \piod}(\gamma_{conv}(ee\hspace {-3.5mm}\nearrow)ee\hspace {-3.5mm}\nearrow)}
\defdecay{\ndalconv} {{\bf\boldmath \pio}(\gamma_{conv}(ee\hspace {-3.5mm}\nearrow)\gamma){\bf\boldmath \piod}(\gamma ee\hspace {-3.5mm}\nearrow)}
\defdecay{\dalcomp} {{\bf\boldmath \pio}(\gamma\gamma){\bf \boldmath\piod}(\gamma_{comp}(e^{-})e^{+}e^{-}\hspace {-1.5mm}\nearrow)}

\defdecay{\pipiconv} {{\bf\boldmath \pio}(\gamma\gamma){\bf\boldmath \pio}(\gamma_{conv}(ee\hspace {-6.5mm}\nearrow)\gamma_{conv}(ee\hspace {-6.5mm}\nearrow))}

\defdecay{\pip} {\pi^+}
\defdecay{\pim} {\pi^-}
\defdecay{\twopic} {\pi^+ \pi^-}
\defdecay{\pipic} {\twopic}
\defdecay{\ktwopic} {\k \rightarrow \twopic}
\defdecay{\kltwopic} {\kl \rightarrow \twopic}
\defdecay{\kstwopic} {\ks \rightarrow \twopic}

\defdecay{\piod} {\pi_{D}^0}
\defdecay{\pioD} {\pi_{Dalitz}^0}
\defdecay{\piodd} {\pi_{DD}^0}
\defdecay{\piopiod} {\pi^0 \pi_{D}^0}
\defdecay{\eeg} {ee\gamma}
\defdecay{\pioeeg} {\pi^0 \rightarrow \eeg}
\defdecay{\kpiopiod} {\k \rightarrow \piopiod}

\defdecay{\kethree} {\mathrm{K_{e3}}}
\defdecay{\pienu} {\pi e \nu}
\defdecay{\klethree} {\kl \rightarrow \pienu}
\defdecay{\kmuthree} {\mathrm{K_{\mu 3}}}
\defdecay{\pimunu} {\pi \mu \nu}
\defdecay{\klmuthree} {\kl \rightarrow \pimunu}
\defdecay{\threepio} {3\pi^0}
\defdecay{\klthreepio} {\kl \rightarrow \threepio}

\defdecay{\Lam}{\Lambda}
\defdecay{\Lambar}{\bar{\Lambda}}
\defdecay{\etagg}{\eta \rightarrow \gamma \gamma}
\defdecay{\etathreepio} {\eta \rightarrow \threepio}
\defdecay{\piogg} {\pi^0 \rightarrow \gamma \gamma}
\defdecay{\meegg} {m_{ee \gamma \gamma}}
\defdecay{\meeggg} {m_{ee \gamma \gamma \gamma}}
\defdecay{\mgg} {m_{\gamma \gamma}}

\defmath{\dvertex}{d_{vertex}}
\defmath{\mgg}{m_{\gamma\gamma}}
\defmath{\chisq}{\chi^2}
\defmath{\rel}{\chi^2}
\defmath{\mone}{m_1}
\defmath{\mtwo}{m_2}

\defmath{\pk}{p_K}
\defmath{\pt}{p_T}
\defmath{\ptp}{{p_T}'}
\defmath{\ptpsq}{{p'_T}^2}
\defmath{\mpp}{m_{\pi\pi}}

\defmath{\asl} {\alpha_{SL}}
\defmath{\asloo} {\alpha^{00}_{SL}}
\defmath{\aslpm} {\alpha^{+-}_{SL}}
\defmath{\Dasl} {\Delta \alpha_{SL}}

\defmath{\als} {\alpha_{LS}}
\defmath{\alsoo} {\alpha^{00}_{LS}}
\defmath{\alspm} {\alpha^{+-}_{LS}}
\defmath{\Dals} {\Delta \alpha_{LS}}

\defmath{\btag} {\beta_{tag}}
\defmath{\btagoo} {\beta^{00}_{tag}}
\defmath{\btagpm} {\beta^{+-}_{tag}}
\defmath{\Dbtag} {\Delta \beta_{tag}}

\defmath{\wpm} {W^{+-}}
\defmath{\woo} {W^{00}}
\defmath{\wooo} {W^{000}}
\defmath{\Dw} {\Delta W}

\defmath{\wt}{W(\tau)}
\defmath{\Dp}{\mathrm{D_p}}

\defmath{\etas}{\eta_S}
\defmath{\etal}{\eta_L}
\defmath{\etasl}{\eta_{S,L}}
\defmath{\lamc}{\lambda^{+-}}
\defmath{\lamn}{\lambda^{00}}
\defmath{\DRint}{(\DR)_{\mbox{\scriptsize intensity}}}
\defmath{\DRgeom}{(\DR)_{\mbox{\scriptsize geometry}}}

\defmath{\R}{R}
\defmath{\DR}{\Delta \R}
\defmath{\epp}{\varepsilon^{\prime}}
\defmath{\vep}{\varepsilon}
\defmath{\epe}{\epp/\vep}
\defmath{\Ree}{\mathcal{R}\!\mathit{e}(\eprime/\epsilon)}
\defmath{\epm}{\eta_{+-}}
\defmath{\eoo}{\eta_{00}}

\defmath{\mum}{\mu\mathrm{m}}
\defmath{\mus}{\mu\mathrm{s}}
\defmath{\degrees}{^{\circ}}
\defmath{\taus}{\tau_S}
\defmath{\taul}{\tau_L}

\defmath{\about}{\sim}
\defmath{\eop}{E/p}
\defmath{\Qx} {Q_x}
\defmath{\twotrack}{2track}
\defmath{\etot}{E_{tot}}
\defmath{\mk}{m_K}
\defmath{\stat}{\mbox{stat}}
\defmath{\syst}{\mbox{syst}}
\defmath{\rcog}{R_{cog}}
\defmath{\mm}{m}
\defmath{\mee}{m_{ee}}
\defmath{\mpi}{m_{\pio}}

\usepackage{amssymb}

\begin{document}
\begin{titlepage}
\docnum{CERN--EP/2003-062}
\date{9~September~2003}
\title{\bf \Large Observation of the rare decay $\kspiee$}
\begin{Authlist}
\begin{center}
%{\bf Experiment NA48/1}
\  \\[0.2cm] 
% A.~Lai,
% D.~Marras \\
%{\em \small Dipartimento di Fisica dell'Universit\`a e Sezione dell'INFN di Cagliari, I-09100 Cagliari, Italy} \\[0.2cm] 
 % 
%
J.R.~Batley,
G.E.~Kalmus\footnotemark[1],
C.~Lazzeroni,
D.J.~Munday,
M.~Patel,
M.W.~Slater,
S.A.~Wotton \\
{\em \small Cavendish Laboratory, University of Cambridge, Cambridge, CB3 0HE,
U.K.\footnotemark[2]} \\[0.2cm] 
 R.~Arcidiacono,
 G.~Bocquet,
 A.~Ceccucci,
 D.~Cundy\footnotemark[3],
 N.~Doble\footnotemark[4],
 V.~Falaleev,
 L.~Gatignon,
 A.~Gonidec,
 P.~Grafstr\"om,
 W.~Kubischta,
F.~Marchetto\footnotemark[5],
 I.~Mikulec\footnotemark[6],
 A.~Norton,
 B.~Panzer-Steindel,
P.~Rubin\footnotemark[7],
 H.~Wahl\footnotemark[8] \\
{\em \small CERN, CH-1211 Gen\`eve 23, Switzerland} \\[0.2cm] 
E.~Goudzovski,
D.~Gurev, 
P.~Hristov\footnotemark[9],
V.~Kekelidze,
L.~Litov,
D.~Madigozhin,
N.~Molokanova,
Yu.~Potrebenikov,
S.~Stoynev,
A.~Zinchenko\\
{\em \small Joint Institute for Nuclear Research, Dubna, Russian    Federation} \\[0.2cm] 
 %no list received!
E.~Monnier\footnotemark[10],
E.~Swallow,
R.~Winston\\
{\em \small The Enrico fermi Institute, The University of Chicago, Chicago, Illinois, 60126, U.S.A.}\\[0.2cm]
 R.~Sacco\footnotemark[11],
 A.~Walker \\
{\em \small Department of Physics and Astronomy, University of    Edinburgh, JCMB King's Buildings, Mayfield Road, Edinburgh,    EH9 3JZ, U.K.} \\[0.2cm] 
 %no list received!
%
W.~Baldini,
P.~Dalpiaz,
P.L.~Frabetti,
A.~Gianoli,
M.~Martini,
F.~Petrucci,
M.~Scarpa,
M.~Savri\'e \\
{\em \small Dipartimento di Fisica dell'Universit\`a e Sezione    dell'INFN di Ferrara, I-44100 Ferrara, Italy} \\[0.2cm] 
 %no list received!
%
%
A.~Bizzeti\footnotemark[12],
M.~Calvetti,
G.~Collazuol,
G.~Graziani,
E.~Iacopini,
M.~Lenti,
F.~Martelli\footnotemark[13],
G.~Ruggiero,
M.~Veltri\footnotemark[13] \\
{\em \small Dipartimento di Fisica dell'Universit\`a e Sezione    dell'INFN di Firenze, I-50125 Firenze, Italy} \\[0.2cm] 
%no list received!
%
%
M.~Behler,
K.~Eppard,
 M.~Eppard,
 A.~Hirstius,
 K.~Kleinknecht,
 U.~Koch,
L.~Masetti, 
P.~Marouelli,
U.~Moosbrugger,
C.~Morales Morales,
 A.~Peters,
 R.~Wanke,
 A.~Winhart \\
{\em \small Institut f\"ur Physik, Universit\"at Mainz, D-55099 Mainz,
Germany\footnotemark[14]} \\[0.2cm] 
A.~Dabrowski,
T.~Fonseca Martin,
S.~Goy Lopez,
M.~Velasco \\
{\em \small Department of Physics and Astronomy, Northwestern University, Evanston Illinois 60208-3112, U.S.A.}
 \\[0.2cm] 
G.~Anzivino,
P.~Cenci,
E.~Imbergamo,
G.~Lamanna,
P.~Lubrano,
A.~Michetti,
A.~Nappi,
M.~Pepe,
M.C.~Petrucci,
M.~Piccini,
M.~Valdata \\
{\em \small Dipartimento di Fisica dell'Universit\`a e Sezione    dell'INFN di Perugia, I-06100 Perugia, Italy} \\[0.2cm] 
 %no list received  !  
% 
%
 C.~Cerri,
F.~Costantini,
 R.~Fantechi,
 L.~Fiorini,
 S.~Giudici,
 I.~Mannelli,
G.~Pierazzini,
 M.~Sozzi \\
{\em \small Dipartimento di Fisica, Scuola Normale Superiore e Sezione dell'INFN di Pisa, I-56100 Pisa, Italy} \\[0.2cm] 
%no list received  !  
% 
%
C.~Cheshkov,
J.B.~Cheze,
 M.~De Beer,
 P.~Debu,
G.~Gouge,
G.~Marel,
E.~Mazzucato,
 B.~Peyaud,
 B.~Vallage \\
{\em \small DSM/DAPNIA - CEA Saclay, F-91191 Gif-sur-Yvette, France} \\[0.2cm] 
M.~Holder,
 A.~Maier,
 M.~Ziolkowski \\
{\em \small Fachbereich Physik, Universit\"at Siegen, D-57068 Siegen,
Germany\footnotemark[15]} \\[0.2cm] 
C.~Biino,
N.~Cartiglia,
M.~Clemencic, 
E.~Menichetti,
N.~Pastrone \\
{\em \small Dipartimento di Fisica Sperimentale dell'Universit\`a e    Sezione dell'INFN di Torino,  I-10125 Torino, Italy} \\[0.2cm] 
 W.~Wislicki,
\\
{\em \small Soltan Institute for Nuclear Studies, Laboratory for High    Energy
Physics,  PL-00-681 Warsaw, Poland\footnotemark[16]} \\[0.2cm] 
H.~Dibon,
M.~Jeitler,
M.~Markytan,
G.~Neuhofer,
L.~Widhalm \\
{\em \small \"Osterreichische Akademie der Wissenschaften, Institut  f\"ur
Hochenergiephysik,  A-10560 Wien, Austria\footnotemark[17]} \\[1cm] 
\vspace{0.5cm}
\it{Accepted for publication in Physics Letters B.}
\rm
\end{center}
\setcounter{footnote}{0}
\footnotetext[1]{Present address: Rutherford Appleton Laboratory,
Chilton, Didcot, OX11 0QX, UK} 
\footnotetext[2]{ Funded by the U.K.    Particle Physics and Astronomy Research Council}
\footnotetext[3]{Present address: Instituto di Cosmogeofisica del CNR di Torino, I-10133 Torino, Italy}
\footnotetext[4]{Also at Dipartimento di Fisica dell'Universit\`a e Sezione dell'INFN di Pisa, I-56100 Pisa, Italy}
\footnotetext[5]{On leave from Sezione dell'INFN di Torino,  I-10125 Torino, Italy}
\footnotetext[6]{ On leave from \"Osterreichische Akademie der Wissenschaften, Institut  f\"ur Hochenergiephysik,  A-1050 Wien, Austria}
\footnotetext[7]{On leave from University of Richmond, Richmond, VA, 23173, 
USA; supported in part by the US NSF under award \#0140230}
\footnotetext[8]{Also at Dipartimento di Fisica dell'Universit\`a e Sezione dell'INFN di Ferrara, I-44100 Ferrara, Italy}
\footnotetext[9]{Present address CERN, CH-1211 Gen\`eve 23, Switzerland}
\footnotetext[10]{Also at Centre de Physique des Particules de Marseille, IN2P3-CNRS, Universit\'e 
de la M\'editerran\'e, Marseille, France}
\footnotetext[11]{Present address Laboratoire de l'Acc\'elerateur Lin\'eaire, IN2P3-CNRS, Universit\'e de Paris-Sud, 91898 Orsay, France}
\footnotetext[12]{ Dipartimento di Fisica dell'Universita' di Modena e Reggio Emilia, via G. Campi 213/A I-41100, Modena, Italy}
\footnotetext[13]{ Istituto di Fisica, Universita' di Urbino, I-61029  Urbino, Italy}
\footnotetext[14]{ Funded by the German Federal Minister for    Research and Technology (BMBF) under contract 7MZ18P(4)-TP2}
\footnotetext[15]{ Funded by the German Federal Minister for Research and Technology (BMBF) under contract 056SI74}
\footnotetext[16]{Supported by the Committee for Scientific Research grants
5P03B10120, SPUB-M/CERN/P03/DZ210/2000 and SPB/CERN/P03/DZ146/2002}
\footnotetext[17]{Funded by the Austrian Ministry for Traffic and 
Research under the 
contract GZ 616.360/2-IV GZ 616.363/2-VIII, 
and by the Fonds f\"ur   Wissenschaft und Forschung FWF Nr.~P08929-PHY}

\end{Authlist}
\end{titlepage}

%\setcounter{footnote}{0}
%\begin{linenumbers} 

\begin{abstract}

A search for the decay $\kspiee$ has been made 
%using
by the NA48/1 experiment at the CERN SPS accelerator. 
Using data collected during 89 days in 2002 
with a high-intensity $\ks$ beam, 
7 events were found with a background of 0.15 events.
The branching fraction 
$BR(\kspiee, \  m_{ee} > 0.165$ GeV/{\it c}$^2$) = $(3.0^{+1.5}_{-1.2} (\text{stat}) 
\pm 0.2 (\text{syst})) \times 10^{-9}$ has been measured.
Using a vector matrix element and a form factor equal to one, the measurement
gives $BR(\kspiee) = (5.8^{+2.9}_{-2.4}) \times 10^{-9}$.

\end{abstract}

%\begin{keyword}

%\PACS 87.53.-j \sep 87.53.Qc
%\end{keyword}
% \end{frontmatter}

% main text
% \begin{linenumbers} 

\vspace{0.2cm}

\section{Introduction}
 
 When not forbidden by CP-conservation, the $\kpiee$ decay can proceed
 via single photon exchange. 
 This is the case for $K_S$ and charged kaons, while the $K_L$ decay
 - barring a  small CP-conserving contribution - is CP-violating. 

The rate of \kspiee induced 
by the electromagnetic interaction was predicted in ref.~\cite{sehgal}
to be BR$(\kspiee) = 5.5 \times 10^{-9}$.
%L.~M. Sehgal to be:
%$\frac{\Gamma(K_1 \rightarrow \pi^0 e^+ e^-)}{\Gamma(K_2 \rightarrow \pi^+
%\pi^-)} = 8 \times 10^{-9}$
%$\Gamma(K_1 \rightarrow \pi^0 e^+ e^-) / \Gamma(K_1 \rightarrow \pi^+
%\pi^-) = 8 \times 10^{-9}$~\cite{sehgal}.

 The theoretical aspects of the decay $\kspiee$ were studied 
to leading order in the chiral expansion 
in ref.~\cite{bib:triangle, bib:prades} and 
the implications of this decay with respect to the 
search for CP-violation in rare kaon decays were 
investigated in ref. ~\cite{Ecker:1987hd} and
re-examined in ref.~\cite{bib:gabbiani}.
Further study beyond leading order was presented in ref.~~\cite{bib:ambrosio},
where the branching fraction for $\kspiee$ was expressed 
 as a function of one parameter $a_S$:

\begin{equation}
{\rm BR}(\kspiee) \: =  \: 5.2 \, \times \, 10^{-9}\; a_{\text {\em S}}^2.
\label{eq:as}
\end{equation}

%$$ BR(KS-->pi0 ee) = 5.2 10-9 a_S^2 $$

For $\klpiee $,  CP-violating contributions can 
 originate from:

\begin{enumerate}
\item[a)]
 $\kz - \kbar$ mixing via a decay of the CP-even 
 component of the $\kl$ ($K_1$)
  into $\piee$.
This indirect CP-violating contribution is related 
to the $\ks$ branching ratio:

\begin{equation}
{\rm BR}(\klpiee) = \frac{\tau_L}{\tau_S}|\epsilon|^2  BR(\kspiee )
\simeq \frac{BR(\kspiee)}{330}
\label{eq:cpeven}
\end{equation}%$$ BR(KL-->pi0 e e ) = \frac{\tau_L}{\tau_S}|\epsilon|^2  BR(KS-->pi0 e e )$$

\item[b)]
direct CP-violating contribution from short distance physics 
via loops sensitive 
to $Im (\lambda_t)=Im (V_{td}V^*_{ts})$. 
%the combination of CKM matrix elements 
%relevant for CP-violation in rare kaon decays.  
\end{enumerate}

The indirect and direct CP-violating contributions 
can interfere and the expression 
for the total CP-violating 
branching ratio of \klpiee can be written as~\cite{bib:ambrosio}:  

\begin{equation}
{\rm BR}(\klpiee)_{\text{CPV}} \times 10^{12} \simeq 
\; 15.3 \, a^2_{\text{\em S}} - 6.8 
\, a_{\text{\em S}} \left(\frac{{Im}(\lambda_t)}{10^{-4}}\right)
+ 2.8 \left(\frac{{Im}(\lambda_t)}{10^{-4}}\right)^2
\label{eq:interference}
\end{equation}

As shown in eq. \ref{eq:interference}, 
the sensitivity to $Im (\lambda_t)$ 
can also come from the 
interference term depending on the value of $a_S$.
% albeit with a sign ambiguity. 
The theoretical predictions for $\kspiee$  
do not provide firm constraints on $Im (\lambda_t)$
and a measurement or a stringent upper limit on $a_S$ is 
necessary to progress further in the understanding of CP-violation 
in the $\klpiee$  decay.

Currently, the upper limit of the $BR(\klpiee)$ is $5.1 \times  10^{-10}
$~\cite{Alavi-Harati:2000sk}.
This together with  the present upper limit 
BR$(\kspiee) <1.4 \times 10^{-7}$~\cite{Lai:2001jf}
gives a bound on $Im (\lambda_t)$~\cite{Isidori:2001nd}, but not  
competitive with respect to other 
constraints obtained from b-physics.

\section{Data-taking}

\subsection{Beam}
The experiment was performed at the CERN SPS accelerator, and used a
400 GeV/{\it c} proton beam impinging on a Be target 
to produce a neutral beam.
%$K_S$ beam.
The spill length was 4.8~s out of a 16.2~s cycle time.
The proton intensity was fairly constant during the spill with a mean of 
$5 \times 10^{10}$  particles per pulse.

\begin{figure}[hbt]
\begin{center}
\mbox{\epsfig{file=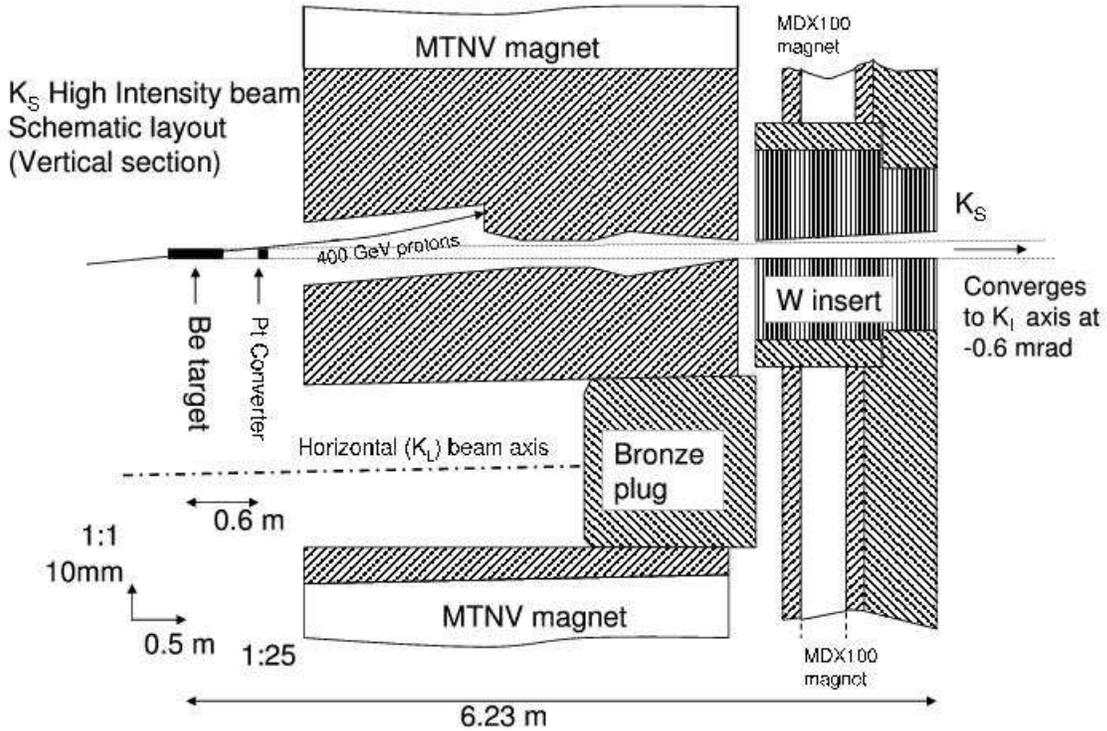,scale=0.6,angle=-90.}}
\caption{View of the 2002 modifications to the beam line.}
\label{beam1}
\end{center}
\end{figure}

Fig. \ref{beam1} shows the modifications with respect to the
previous \ks~ beam line described in ~\cite{bib:epsi}. 
%The $K_S$
%beam intensity was about 3 orders of magnitude greater than that used
%in the $Re(\epsilon^{\prime} / \epsilon)$ measurement~\cite{bib:epsi}.
% An absorber plug was used to block the $\kl$ beam line.
The \kl~ beam line was blocked and
an additional sweeping magnet was installed to cover the defining 
section of the \ks~ collimator.
%To reduce the $K_S$ beam contamination by photons, 
To reduce the number of photons in the neutral beam,
primarily from
$\pi^0$ decays, a platinum absorber 24 mm thick was placed in the 
beam between the target and a sweeping magnet, which deflected charged 
particles.
A 5.1~m thick collimator, the axis of which formed an angle of 4.2~mrad
to the proton beam direction, selected a beam of 
neutral long-lived particles (\KS, \KL,
$\Lambda^0$, $\Xi^0$, $n$ and $\gamma$). 
On average $2 \times 10^{5}$ \KS \ per spill decayed in
the fiducial volume downstream of the collimator with 
a mean energy of 120 GeV.

\subsection{Detector}

The detector was designed for the measurement of $Re(\epsilon^{\prime}
/ \epsilon)$\cite{bib:epsi}.
In order to minimize the interactions of the neutral beam with air,  
the collimator was immediately 
followed by a $\sim $ 90 m long evacuated tank
which was terminated by a 0.3\% $X_0$ thick Kevlar window.
The detector was located downstream  of this tank.
%and is located 114 m downstream of the
%final collimator.

\subsubsection{Tracking}

The detector included a spectrometer 
housed in a helium gas volume
with two drift chambers before and two after
a dipole magnet with a horizontal transverse momentum kick
of 265 MeV/{\it c}. 
%It was placed 114 meters downstream of the collimator ends
Each chamber had four views ({\it x, y, u, v}), each of which had two 
sense wire planes.
The resulting space points were typically reconstructed with a
resolution of $\sim 150$~$\mu$m in each projection.
The spectrometer momentum resolution could be parameterized as:

\begin{equation*}
\sigma_p /p = 0.48 \% \oplus 0.015\% \times p 
\end{equation*}

\noindent where $p$ is in GeV/{\it c}. This
% which 
gave a resolution of 3 MeV/{\it c}$^2$ 
when reconstructing the kaon mass in \kspipi\
decays. The track time resolution was $\sim~1.4$~ns.

\subsubsection{Electromagnetic Calorimetry}

The detection and measurement of the electromagnetic showers 
were achieved with a
liquid krypton calorimeter (LKr), 27 radiation lengths deep,  with a 
$\sim $ 2 cm $\times $ 2 cm cell cross-section.

The energy resolution, expressing $E$ in GeV, 
may be parameterized as \cite{bib:unal}:

\begin{equation*}
\sigma(E) / E = 3.2\% /\sqrt{E} \oplus 9 \% / E \oplus 0.42 \%
\end{equation*}

The transverse position resolution for a single photon
of energy larger than 20 GeV was better than 1.3 mm, and
the corresponding mass resolution at the $\pi^0$ mass was
$\sim$ 1 MeV/{\it c}$^2$.
The time resolution of the calorimeter for a single shower was
better than $\sim~300$~ps.
%and the time of the event could be determined with an accuracy of
%better than 220~ps.   

\subsubsection{Scintillator Detectors}

A scintillator hodoscope was located between the
spectrometer and the calorimeter. It consisted of two planes, segmented
in horizontal and vertical strips and arranged in four quadrants. 
%it was used in the trigger for charged events.
%Each plane consisted of 64 strips, with each strip the length of half
%the plane. 
%The hodoscope provided a resolution of 350~ps for the trigger for charged
%events.
Further downstream there was an iron-scintillator sandwich 
hadron calorimeter, followed by muon counters consisting of three planes 
of scintillator, each shielded by an iron wall.
The fiducial volume of the experiment was principally determined by the 
LKr calorimeter acceptance, together with 
seven rings of scintillation counters 
%which were placed so as 
used to veto activity outside this region.

\subsubsection{Trigger and Readout}

The detector was sampled every 25~ns with no dead time
and the samples were recorded
in a time window of $ 200$~ns encompassing the event trigger time. This
allowed the rate of accidental activity to be investigated in
appropriate time sidebands.

%Data from the whole apparatus was recorded in a time
%window of $\pm 100$~ns around the event trigger.

The event trigger for the signal \kspiee\ had both hardware and
software parts:
\begin{itemize}
\item The hardware trigger \cite{bib:trigneut}
selected events satisfying the following conditions:
  \begin{itemize}
  \item[-] hit multiplicity in the first drift chamber compatible with one or
           more tracks;
  \item[-] hadron calorimeter energy less than 15~GeV;
  \item[-] electromagnetic calorimeter energy greater than 30~GeV;
  \item[-] the centre of energy of the electromagnetic clusters (see eq.~
  \ref{eq:cog} below) less than 15 cm from the beam axis;
  \item[-] the decay occurring within six $K_S$ lifetimes
           from the end of the collimator;
  \item[-] no hits in the two ring scintillator counters farthest 
downstream.
  \end{itemize}
\vspace{0.2cm}
\item The software trigger required:
  \begin{itemize} 
  \item[-] at least two
  tracks in the drift chambers and two extra, well-separated
  clusters each with energy greater than 2 GeV;
  \item[-] the tracks projected from the drift chamber, after the magnet,
  had to match to clusters in the LKr within 5~cm;
  \item[-] the tracks had to be compatible
  with being electrons or positrons using the condition that the ratio
  $E/p$, between the cluster energy in the LKr, $E$, and the momentum
  measured with the drift chambers, $p$,
  had to be greater than 0.85. 
  \item[-] a cluster separation of more
  than 5~cm was required to limit the degradation of the energy
  resolution due to energy sharing between closely spaced clusters. 
  \end{itemize}
\end{itemize}

The events that satisfied the trigger conditions were recorded and
reprocessed with improved calibrations to obtain the final data sample.
 
\subsection{Event selection}

For the analysis of the data, signal and control regions
were defined. These regions were masked while the cuts to reject the
background were tuned using both data and Monte Carlo simulation.

The signal channel \kspiee\ required the identification of an
electron and a positron accompanied by two additional clusters in the LKr.

Tracks reconstructed from the spectrometer
which matched an LKr cluster were labelled as an {\it electron} or {\it 
positron} by requiring three conditions to be met:
no more than 3~ns difference between track time and cluster time;
$0.95 < E/p < 1.05$; and less than 2~cm between the projected track and the
cluster coordinates in the LKr.

We define $\Delta t$ to be the difference between the
average time of the two clusters associated with tracks and the
average time of the two neutral clusters. 
Events were accepted if $ | \Delta t| < 3 $ ns.
 
Events with extra tracks or extra clusters
within 3 ns of the average time of the tracks or clusters 
and with an energy larger than 1.5~GeV were rejected.
To minimize the
effect of energy sharing on cluster reconstruction,
a minimum cluster separation of 10~cm was imposed.
In addition, a distance greater than 
2~cm between the impact points of the two tracks
at the first drift chamber was required.

Four quantities related to the decay vertex were computed.
\begin{itemize}
\item {\it neutral vertex}.

\noindent The vertex position was computed 
from the energies and positions of the four clusters
in the LKr according to

\begin{equation}
z_{\text{\em neutral}} = z_{LKr} - \sqrt{\Sigma_{i,j>i} E_i E_j d_{ij}^2} / M_K
\end{equation}

%\vskip 0.5cm
\noindent where $z_{LKr}$ is the longitudinal position of the front
face of the LKr;
%\noindent where $d_{LKR}$ is the vertex distance from the calorimeter,
$M_K$ is the kaon mass, 
$E_{i,j}$ is the energy of the $(i,j)^{th}$ cluster and $d_{i,j}$ 
is the distance between clusters $i$ and $j$. 
In the case of
 the photons these are the $x,y$ shower positions in the LKr. 
For the $e^{\pm}$ tracks,
 in order to cancel the deflection due to the dipole magnet, the ({\it x,y}) 
 positions were calculated by extrapolating
 the tracks from their positions in the first two drift chambers to the face
of the LKr.
The $x$ and $y$ coordinates of the neutral vertex were found by extrapolating
the position of each track before the magnet to the position
of $z_{\text{\em neutral}}$. The average of the two measurements was
taken as the $(x,y)$ vertex position.

The neutral vertex was used to compute 
the invariant mass of the two photons, $\mgg$. 
% which are
%the (x,y) positions of the $e^+$ and $e^-$ tracks extrapolated to the face
%of the LKR, using their positions at 
%the two drift chambers before the magnet.

\item {\it charged vertex}.

The 
%$z$ 
position of the charged vertex can be calculated
using the constraint that the kaon decay should lie on the straight line 
joining the target and the point defined as $(x_{cog},y_{cog})$: 

\begin{equation}
x_{cog}={{(\sum_i E_i x_i)}}/{\sum_i E_i} ~~~~
 y_{cog}= {{(\sum_i E_i y_i)}}/{\sum_i E_i}
\label{eq:cog}
\end{equation}

\noindent where $E_{i}$, $x_{i}$ and $y_{i}$ are the energy and
positions of the $i$-th cluster. 

For each track, the closest distance of approach between this line and
the track was found, giving two measurements 
%of the z position 
which were then averaged to give the charged vertex position.
% along the beam direction.

The charged vertex was then used to compute $\meegg$, the invariant mass
of the four decay products. 

\item {\it $\pi^0$ vertex}.

The $\pi^0$ vertex position along the beam direction  
was computed in a similar way to the neutral
vertex, but using only the two photon clusters and imposing the
$\pi^0$ mass, $M_{\pi^0}$, instead of the kaon mass.

\item {\it track vertex}.

The track vertex is at the position of the closest 
distance of approach of the two tracks.
\end{itemize}

The $z$ position of the $\pi^0$ and track vertices
had to be greater than
50 cm (one standard deviation) 
%\footnote{ 50 cm corresponds to one standard deviation.}
beyond the collimator exit in order to reject any
interactions occurring in the collimator. 
Assuming the observed event to be a kaon decay,
the proper lifetime was computed from the position of the neutral vertex,
taking the end of the final collimator as the origin.
A cut at 2.5 $K_S$ lifetimes was then applied. The kaon momentum was
required to be between 40 and 240~GeV/{\it c}.

\section{Signal and Control regions}

%Candidate events are expected to have $m_{\gamma
%  \gamma}\approx M_{\pi^0}$ and $m_{e^{+}e^{-}\gamma
%  \gamma}\approx M_{K}$.

The signal region was defined as :
\begin{itemize}
\item $| m_{\gamma \gamma} - M_{\pi^0}| < 2.5 \times
  \sigma_{\mgg}$
\item $| \meegg - M_K| < 2.5 \times \sigma_{\meegg}$
\end{itemize}

To evaluate the resolutions, $\sigma_{\meegg}$ and $\sigma_{\mgg}$,
we studied the channel \kspipid \footnote{ $\pi^0_D$ is
the Dalitz decay $\pi^0 \rightarrow e^+ e^- \gamma$.}, for which we measured 
$\sigma_{\meeggg}$ = 6.5 MeV/{\it c}$^2$ and $\sigma_{\mgg}$ = 1 MeV/{\it c}$^2$
respectively.
These values were found to be 
in agreement with a Monte Carlo simulation
based on GEANT \cite{bib:geant}.
%The $\meeggg$ resolution, $\sigma_{\meeggg}$,
%was measured to be 6.5 MeV/{\it c}$^2$ for the channel \kspipid, 
%in agreement with the Monte Carlo expectation. 
For the decay \kspiee, the Monte Carlo prediction of $\sigma_{\meegg}$ 
was 4.6 MeV/{\it c}$^2$ and this value was used in defining the
signal region.
The better resolution is due to the fact that the $e^+ e^-$ opening
angle 
%for $m_{ee} > 165$ MeV/{\it c}$^2$ 
is on average larger than for
the decay \kspipid. 

The $\mgg$ resolution, $\sigma_{\mgg}$, at the
${\pi^0}$ mass was found to be 1 MeV/{\it c}$^2$ in agreement
with the Monte Carlo simulation.

A control region was also defined as:
\begin{itemize}
\item  $3 \times \sigma_{\mgg} < | \mgg - M_{\pi^0}| < 6 
\times \sigma_{\mgg}$
\item $3 \times \sigma_{\meegg} < | \meegg - M_K| < 6 \times \sigma_{\meegg}$ 
\end{itemize}

Both the signal and the control regions were kept masked while cuts to
reject the background were studied.

\section{Background Rejection}

A large number of possible background channels was studied. These
channels were of two types: 

\begin{itemize}
\item a single kaon or
hyperon decay which reproduced an event falling into the \kspiee\ signal
region% eg $\Xi^0 \rightarrow \Lambda \pi^0$ where $\Lambda \rightarrow
%p \pi^-$ and the proton and $\pi^-$ deposit virtually all their energy
%in the LKR  
\item fragments from two primary decays which
  happen to coincide in time and space and fall into the signal box.
%  As an example we mention the  
\end{itemize} 

%The principle backgrounds are summarized in Table~\ref{backphy}.  

The background contribution from the channels considered was reduced
by imposing additional requirements.
%%%%%%%%

A background source is from the decay $\kspiopio$ where two 
photons from different $\pi^0$'s   
converted either internally (i.e. $\kspidpid$) or externally
and one electron and one positron from different $\pi^0$'s
were outside the detector acceptance.
In order to reject events from this source 
the invariant
masses of the two electron-photon pairs,
$m_{e^{+}\gamma_{1}}$,$m_{e^{-}\gamma_{2}}$ and
$m_{e^{+}\gamma_{2}}$,$m_{e^{-}\gamma_{1}}$, were computed 
using the charged vertex position.
A priori, the combination corresponding to electron-photon 
from the same $\pi^0$ has an invariant mass
smaller than $M_{\pi^0}$.
Thus events were rejected if both $m_{e \gamma}$ masses were
measured to be smaller than  $M_{\pi^0}+ \delta $.
The constant $\delta$ was chosen equal to
30 MeV/c$^2$, which corresponded to $\sim 10 ~\sigma_{m_{e \gamma}}$. 

%This value safely excluded the tail due to measurement
%errors.

\begin{figure}[hbtp]
  \vspace{9pt}
  \centerline{\hbox{ \hspace{0.0in} 
    \epsfxsize=3.0in
    \epsffile{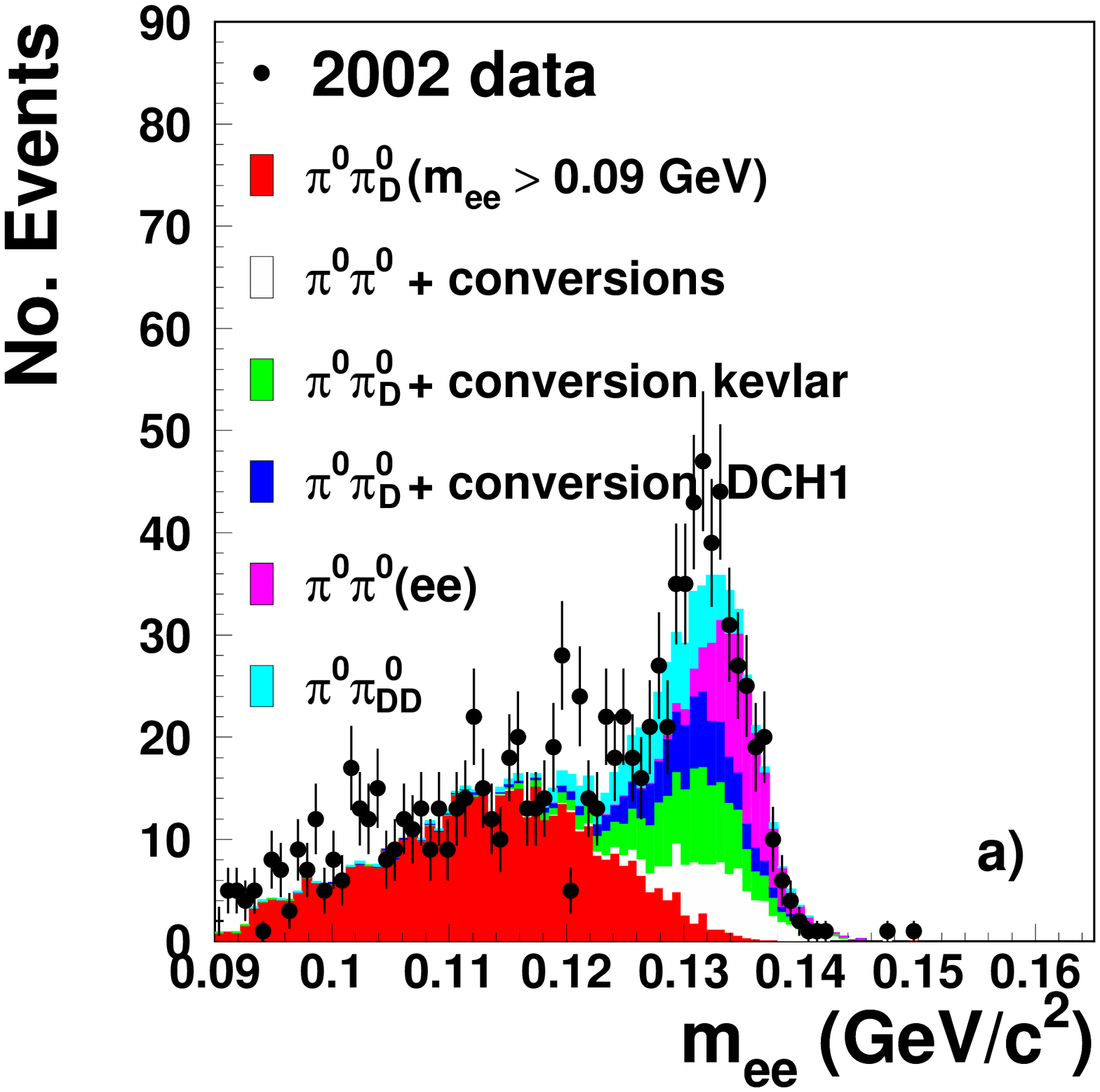}
    \hspace{0.25in}
    \epsfxsize=3.0in
    \epsffile{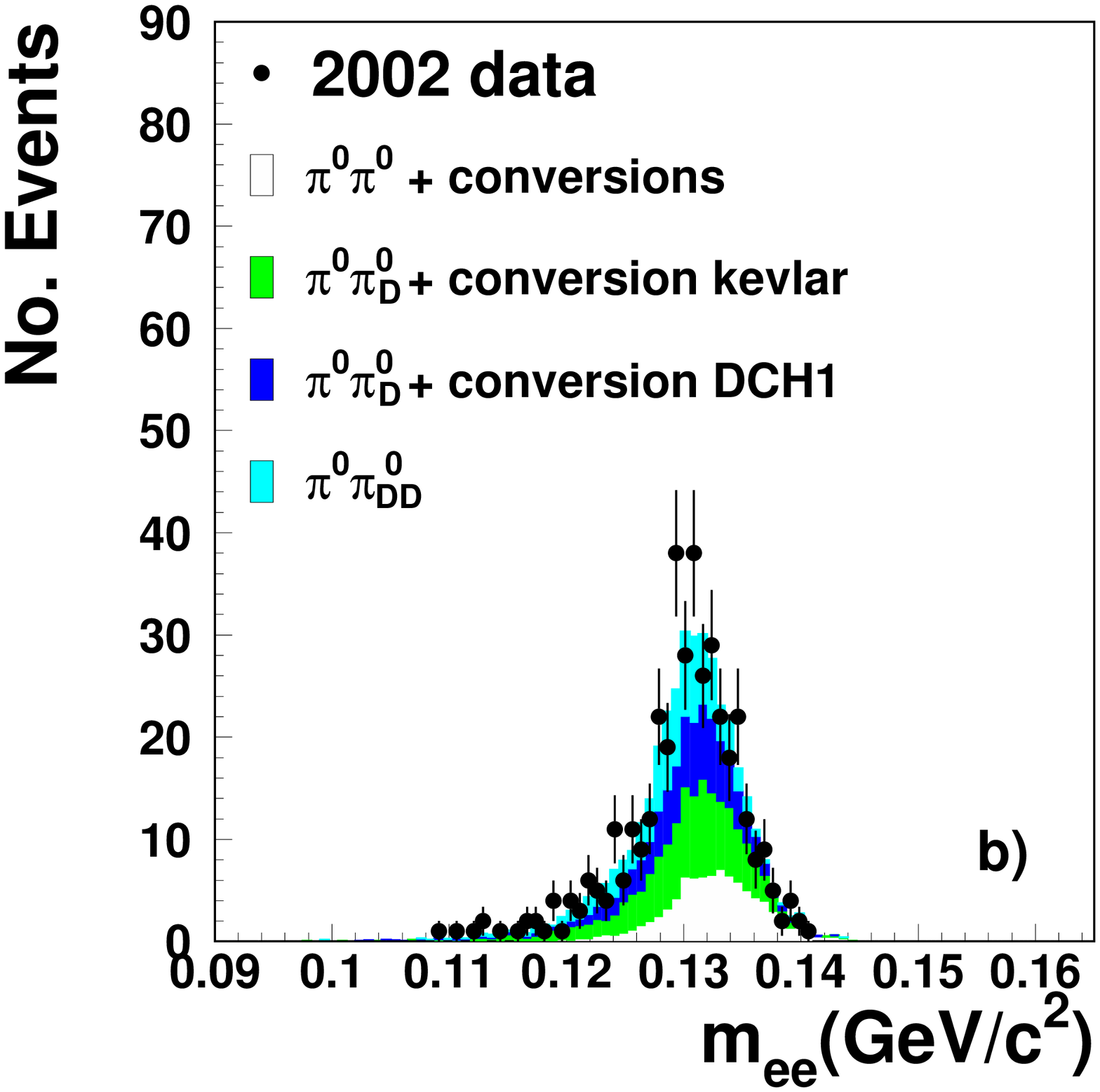}
    }
  }
%  \vspace{9pt}
%  \hbox{\hspace{1.35in} (a) \hspace{2.10in} (b)} 
%  \vspace{9pt}
  \caption{Distributions of $m_{ee}$ after all the cuts have been applied.
Superimposed we show the Monte Carlo predictions from
all important sources. Figure a) shows the components with
opposite-sign tracks; figure b) shows the ones with same-sign tracks.}
  \label{mee}
\end{figure}

Another source of background was due to 
$\pi^0\pi^0$ decays where one
or more photons from a single $\pi^0$ decay converted (either
internally or externally).
These decays are kinematically constrained to have $m_{e^+e^-}<  M_{\pi^0}$ and
in order to reject this background the analysis was restricted to 
the event sample with invariant mass $m_{e^+
e^-} > M_{\pi^0}+ \epsilon $.
%Also in this case
%$\delta$ was determined to safely exclude up to 
%the tail due to measurement
%errors.
To determine $\epsilon$, we analysed
the $m_{ee}$ distribution from data and 
compared it to a Monte Carlo
simulation
where the different components were identified.
In fig.~\ref{mee}.a we show the $m_{ee}$ distribution 
for data (full dots) and superimposed  
the contributions from of all relevant background 
sources. Above the $\pi^0$ mass
the tail of the $m_{ee}$ distribution falls rapidly to zero.
The constant $\epsilon$ was also chosen equal to 30 MeV/{\it c}$^2$ and
the analysis was therefore restricted to the region
$m_{e^+ e^-} >  M_{\pi^0} + 30$ MeV/c$^2$ = 165 MeV/c$^2$,
where $\gamma$ conversions or decays from a single $\pi^0$ give
a negligible
contribution to the background.
This was confirmed from the analysis of events 
with same sign-tracks. This sample
contained events where both photons from a single $\pi^0$
converted and both the electrons or the positrons were 
in the acceptance.
The $m_{e^{\pm} e^{\pm}}$ distribution is shown in
fig.~\ref{mee}.b, where data and Monte Carlo are compared.
No events with $m_{e^{\pm} e^{\pm}} > 165$ MeV/c$^2$ were found. 

%decays in which an
%$e^+ e^- $ are selected from a single \pio\ decay make a negligible
%contribution to the background. 
   
%In fig.~\ref{mee} a) we show the $m_{ee}$ distribution 
%for opposite-sign tracks, 
%while in fig. \ref{mee} b) it is shown the same-sign tracks.
%By subtracting the same and opposite sign distributions, the components
%from \kspipid\ and $\pi^0 \pi^0 (ee)$ were isolated (Fig. \ref{pi0ee}).  

%The $m_{e^+ e^-}$ resolution around the $\pi^0$ mass was then determined 
%to be 3 MeV/$c^2$. 

%The background from \kspidpid\ decays, where one soft $e^{\pm}$ is
%lost from each $\pi^0_D$ was rejected using a cut on the invariant
%masses $m_{e^{+}\gamma_{1}}$,$m_{e^{-}\gamma_{2}}$ and
%$m_{e^{+}\gamma_{2}}$,$m_{e^{-}\gamma_{1}}$ of electron-photon
%pairs. For each of the four mass combinations it was required that at
%least one combination had $m_{e\gamma}>(M_{\pi^0} + 30$ MeV/{\it
%  c}$^2$). This requirement also rejects $\pi^{0}{\pi^{0}}_D$ events in
%which one of the photons from the non-Dalitz decaying $\pi^{0}$
%converts externally.  
  
To reject the background due to electron bremsstrahlung,
the invariant mass of any $e \gamma$ combination was required to be
larger than 20~MeV/{\it c}$^2$.

The background from $\Xi^0 \rightarrow \Lambda \pi^0$ and $\Lambda
\rightarrow p \pi^-$ decays was reduced to a negligible level by
exploiting the large momentum asymmetry in both the $\Lambda \pi^0$
and the $p \pi^-$ final states. \kspiee\ candidates were required to
have $(P_{\Lambda} - P_{\pi^0})/ (P_{\Lambda} + P_{\pi^0})$  
smaller than 0.4 or $(P_p - P_{\pi^-})/(P_p + P_{\pi^-})$ 
smaller than 0.5.
A similar cut was used to remove  \cascadebar~and \lambdabar. 

The possibility of proton and pion misidentification as $e^{\pm}$ was
considered and final states which contained these particles were found
to make a negligible contribution to the background after the
application of the $E/p$ requirement.

%The $m_{\gamma \gamma}$ versus $m_{ee} $
%scatter plot shown in Fig. \ref{mggmee} b) was obtained after the 
%application of these cuts.

%Another potentially dangerous background arises from the following
%decay chain:
%$\Xi^0 \rightarrow \Lambda \pi^0$ and $\Lambda \rightarrow p \pi^-$, 
%where both the proton and the $\pi^-$ shower in the
%LKR and the reconstructed energies satisfy the $E/p$ condition.
%This has a small but not negligible probability.
%These events were characterized by large momentum asymmetry both 
%for the $\Xi^0 \rightarrow \Lambda \pi^0$ and 
%for the $\Lambda \rightarrow p \pi^-$.
%This background was completely 
%rejected by requiring that $|P_{\Lambda} - P_{\pi^0}|/
%(P_{\Lambda} + P_{\pi^0})$ be smaller than 0.4 or $|P_p - P_{\pi^-}|/
%(P_p + P_{\pi^-})$ to be smaller than 0.5.

% The plane $m_{\gamma \gamma}$ versus
% $m_K$ was examined, where $m_K$ is the total invariant mass reconstructed 
% assuming the decay products come from the charged vertex.

%The $\sigma_{m_K}$ for the decay \kspiee was predicted by Monte Carlo
%to be $\sigma_{m_K} = 4.6 MeV/{\it c}^2$, less than the one measured
%using the channel \kspipid \ since \kspiee , for
%$m_{ee} > 165$ MeV/{\it c}$^2$, has a larger opening angles
%between the two tracks.
%

\section{Estimate of the Residual Background}

\begin{figure}[hbt]
\begin{center}
\mbox{\epsfig{file=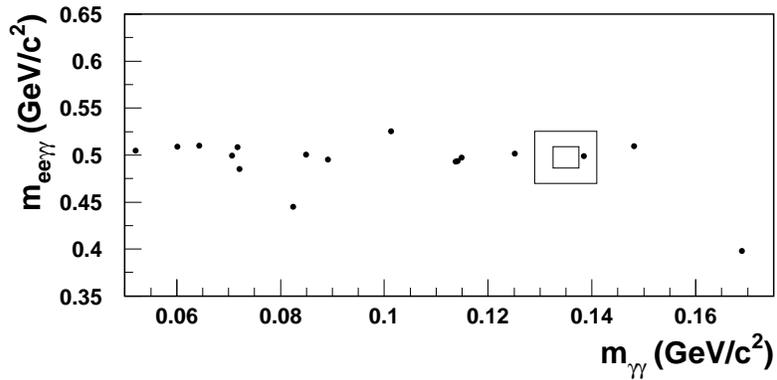,width=12cm}}
\caption{Scatter plot of $\meegg$ versus $m_{\gamma \gamma}$ for events
selected as $K_L \rightarrow e^+ e^- \gamma \gamma$ in the 2001 data.
The boxes are representing the $3 \sigma$ and $6 \sigma$ regions.}
\label{eegg}
\end{center}
\end{figure}

After the selection outlined above three sources of
background were found to be non-negligible: 

\begin{itemize} 

%\item {\it Backgrounds from particle decays}. 
%  \begin{itemize} 
  \item[1.] 
%the irreducible background from the channels
  $K_{L,S} \rightarrow e^+ e^- \gamma \gamma$ 
  %(dominated by $K_L$). 
  \par\noindent The $e^+ e^- \gamma \gamma$
  component was measured using $K_L$ data from the 2001 
  %$K_L$ and $K_S$
  run, in which the 
  number of 
  $K_L\rightarrow e^+ e^- \gamma \gamma$ decays 
  was $\sim$ 10 times the 
  sum of the \kl~and $K_S\rightarrow e^+ e^- \gamma \gamma$ 
  expected in the
  present experiment.
  The distribution of $\meegg$
  versus $\mgg$ for these events is shown in
  fig.~\ref{eegg}.  
  Using a linear extrapolation from the low
  $\mgg$ region to the signal region,
  the background from this channel was estimated to be
  $0.08^{+0.03} _{-0.02}$ events. 
 
  \item[2.] \kspidpid\ 
%  which mimic the $\pi^0 e^+ e^-$ signal due to the loss
%  of one or more particles outside the experimental acceptance.
%  The contribution from this category was mainly $\pi^0_{D} \pi^0_{D}$ 
%  decays with an electron and a positron from different $\pi^0$s
%  swept out by the magnet.
  
  \par\noindent  This was evaluated using full Monte Carlo 
  simulation for a sample which was
  30 times greater than the data,
  and the background was 
  estimated to be less than 0.01 events in the signal region.

%  \end{itemize}

\item[3.] {\it Accidental backgrounds}.

%begin{itemize} 
%  \item[3.] $K_{L,S} \rightarrow  \pi^{\pm} e^{\mp} \nu$ and $\pio$ 

\par\noindent This component was studied using data with the timing
requirements relaxed. Events in the time sidebands,
satisfying all the other cuts, were used to 
extrapolate the background from the control to the signal region.
A further correction was applied to account for the background shape 
in the $m_{\gamma \gamma}$ versus $m_{e^+ e^- \gamma\gamma}$ plane as predicted by a
simulation. 
\par\noindent

%The distributions of $m_{e^+ e^- \gamma\gamma}$ and $m_{\gamma\gamma}$
%versus $\Delta t$ for events in the time sidebands are shown in
%fig.~\ref{aot}.
 %$\Delta t$ is defined as the difference between the
%average time of the two clusters associated with tracks and the
%average time of the two neutral clusters. 
\par\noindent The contribution due to this component was
$0.07^{+0.07}_{-0.03}$ events in the signal region.
\end{itemize}

Other sources of background were considered, for instance that
due to resonances produced by a single proton in the target, and
decaying to a pair of kaons or a $K \Lambda$ pair in the fiducial
region. These contributions were found to be negligible.

%These could mimic a $\pi^0 e^+ e^-$ final
%state if fragments of each decay are lost and there is 
%hadron to electron mis-identification.
%This contribution was found
%to be negligible.

With all the cuts applied, the control region 
was unmasked to estimate the final background contribution to the signal.
No events were found in the control region, consistent with the
background prediction of 0.33 events.
Only one background event
was found in a much larger region (corresponding to $17 \times \sigma_{m_K}$ 
and $20 \times \sigma_{m_{\gamma \gamma}}$).
The background estimate is summarised in Table \ref{back}.
\par The resulting estimate of the total background in the signal
region was $0.15^{+0.10} _{-0.04}$ events.

\section{Normalisation.}

The trigger efficiency was measured using
a control sample of $\sim 8.6 \times 10^6$ \kspipid \ decays, 
which differed topologically
from \kspiee \ only in having an extra photon.
This sample was collected with the same trigger chain.
The trigger efficiency,
measured with a sample of triggers 
collected requiring minimal bias conditions, 
was found to be 99.0\%.  
%The Monte Carlo simulation was checked using the
%$\pi^0 \pi^0_D$ sample,
%and was found to be in good agreement with the data.
The acceptance, including the selection criteria, was found to be
3.3\% for $\pi^0 \pi^0_D$, evaluated using a Monte Carlo simulation.
The Monte Carlo simulation was found to be in good
agreement with the \kspipid\ data.
%in which the radiative corrections for final state interactions
%were taken into account~\cite{bib:photos}.
To obtain the \kspiee~branching ratio, the $K_S$ flux
was calculated using the channel \kspipid \ for normalisation, which 
was selected using the same trigger.
Using the value for the 
branching ratio $BR(\kspipid) = 7.43 \times 10^{-3}$~\cite{bib:PDG}
the $K_S$ flux was calculated,
%of $3.28 \pm 0.16 \times 10^{10}$,
for kaon energies between 40 and 240 GeV/{\it c} 
and kaon lifetimes between zero and 2.5 $K_S$ mean lifetimes
from the collimator exit.
%The $K_S$ flux was also calculated from the $K_S \rightarrow \pi^0 \pi^0$ 
%channel, and was found to be 18.8\% higher.
%Part of the difference could be explained by inefficiencies in the
%drift chamber readout at the start of the run and by accidental losses
%which affected the $\pi^0 \pi^0_D$ trigger but not the $\pi^0 \pi^0$ one.
%To take into account the remaining difference of about 10\%,
%the $\pi^0 \pi^0_D$ flux was increased by 5\% and 
%a systematic uncertainty of 5\% was assigned to the measurement.
The total number of $\ks$ decaying 
%in the momentum range $40$ to $240$ GeV/{\it c} 
within the fiducial volume was $(3.51 \pm 0.17) \times 10^{10}$.

\section{Result}

\begin{figure}[hbtp]
  \vspace{9pt}
  \centerline{\hbox{ \hspace{0.0in} 
    \epsfxsize=3.0in
    \epsffile{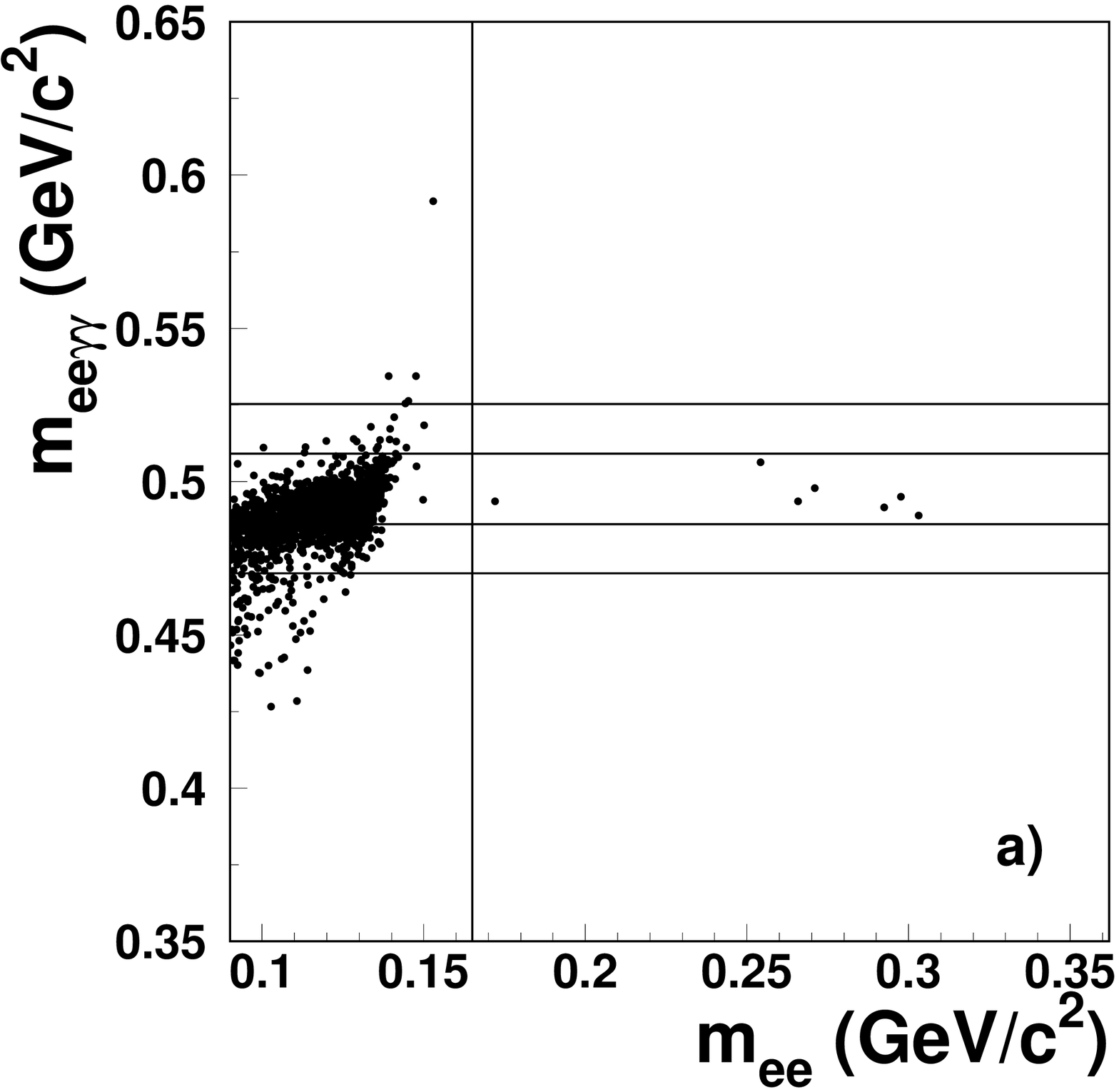}
    \hspace{0.25in}
    \epsfxsize=3.0in
    \epsffile{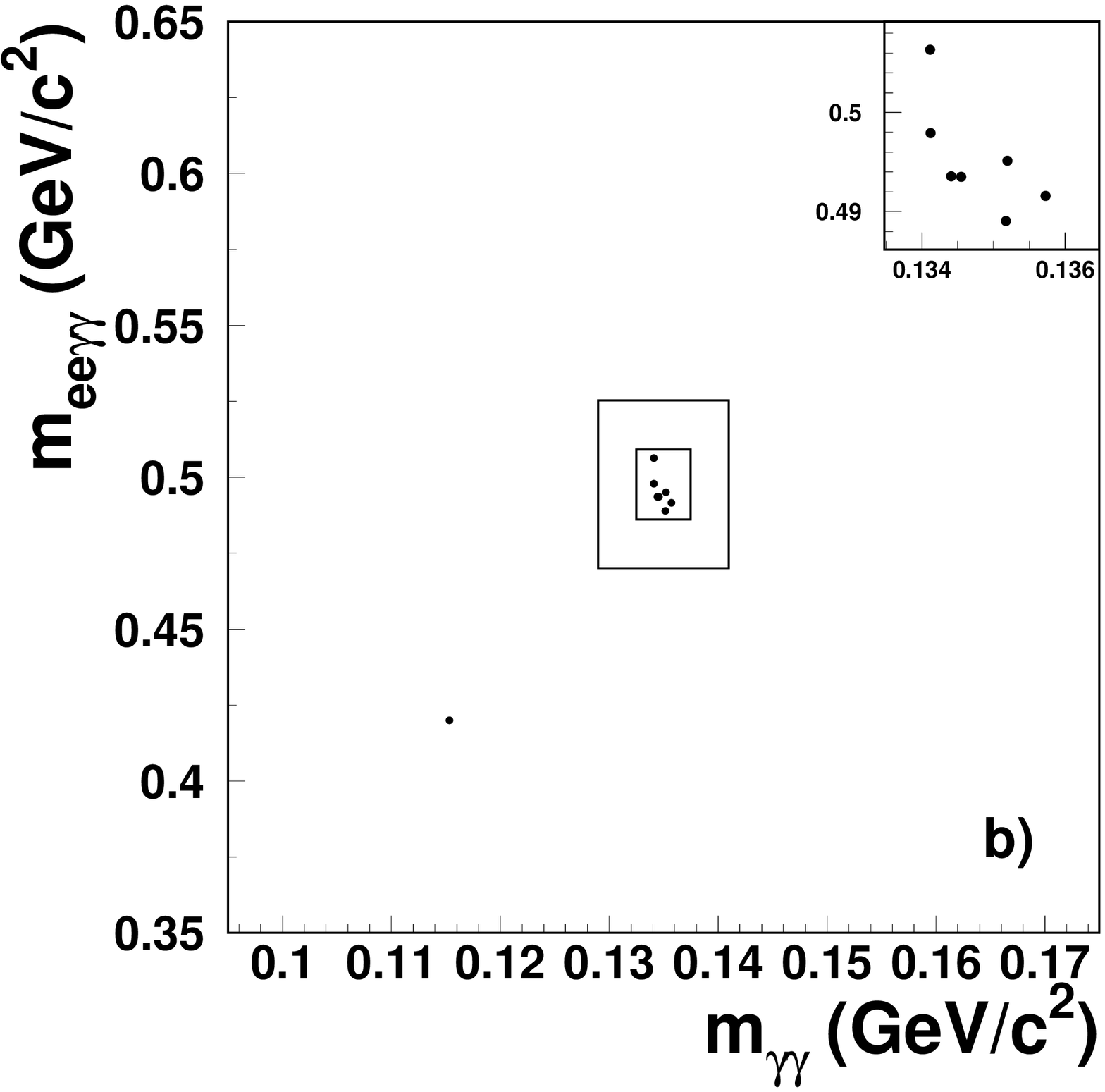}
    }
  }
%  \vspace{9pt}
%  \hbox{\hspace{1.35in} (a) \hspace{2.10in} (b)} 
%  \vspace{9pt}
\caption{Scatter plot of  $\meegg$  versus $m_{ee}$(a) and
$\meegg$ versus $m_{\gamma \gamma}$ (b) for events
passing all the cuts described in the text.
 The regions of $3 \sigma$ and $6 \sigma$ are shown.}
\label{mggmk}
\end{figure}

When the signal region was unmasked seven events were found  
(fig.~\ref{mggmk}).
With an expected background of $0.15^{+0.10} _{-0.04}$ events, 
this corresponds
to a signal of $6.85^{+3.8}_{-1.8}$.
The probability that such a signal is consistent
with background 
is $\sim 10^{-10}$.
We therefore interpret the
signal as the first observation of the $\kspiee$ decays.

\begin{figure}[hbtp]
  \vspace{9pt}
  \centerline{\hbox{ \hspace{0.0in} 
    \epsfxsize=3.0in
    \epsffile{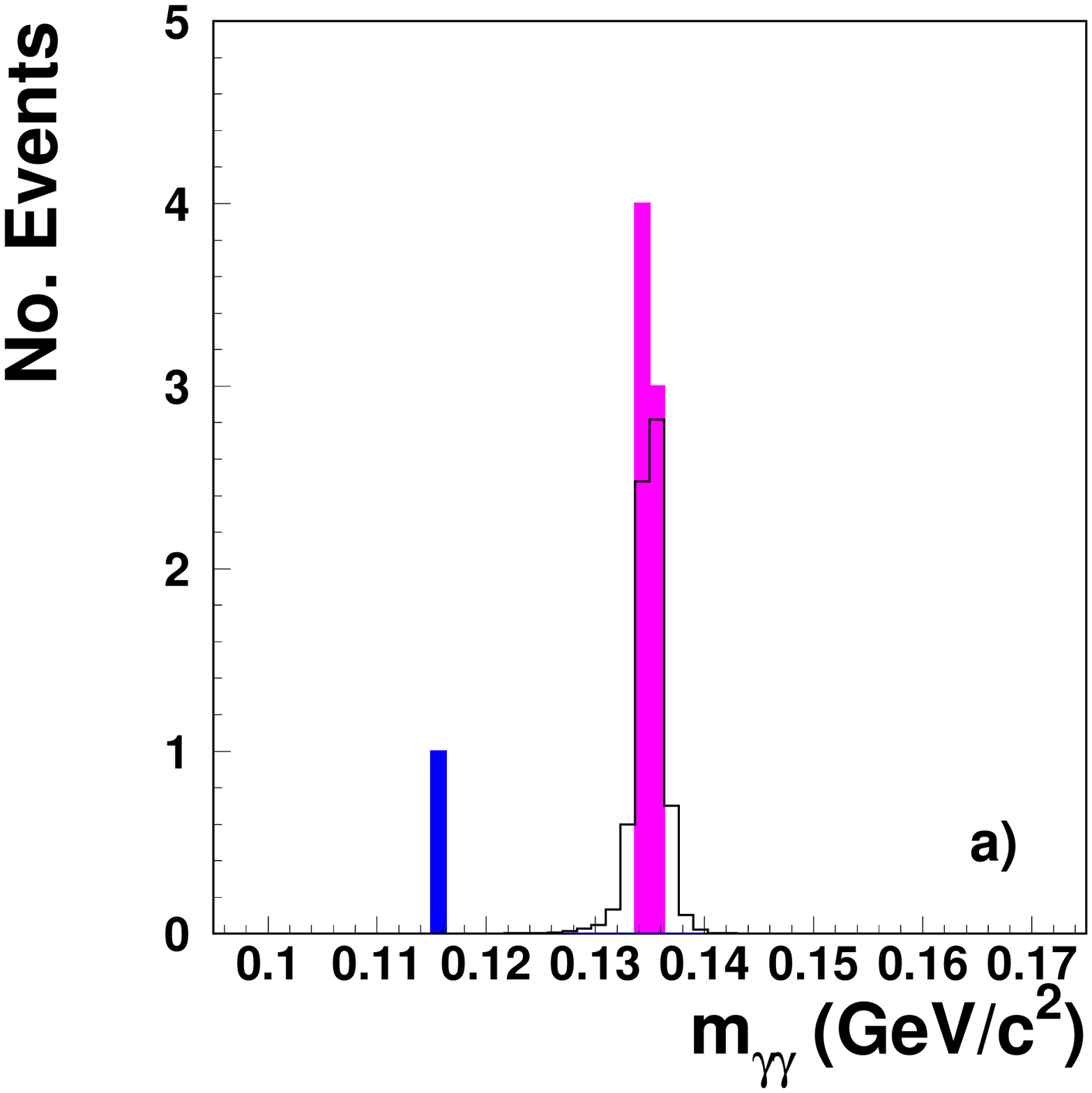}
    \hspace{0.25in}
    \epsfxsize=3.0in
    \epsffile{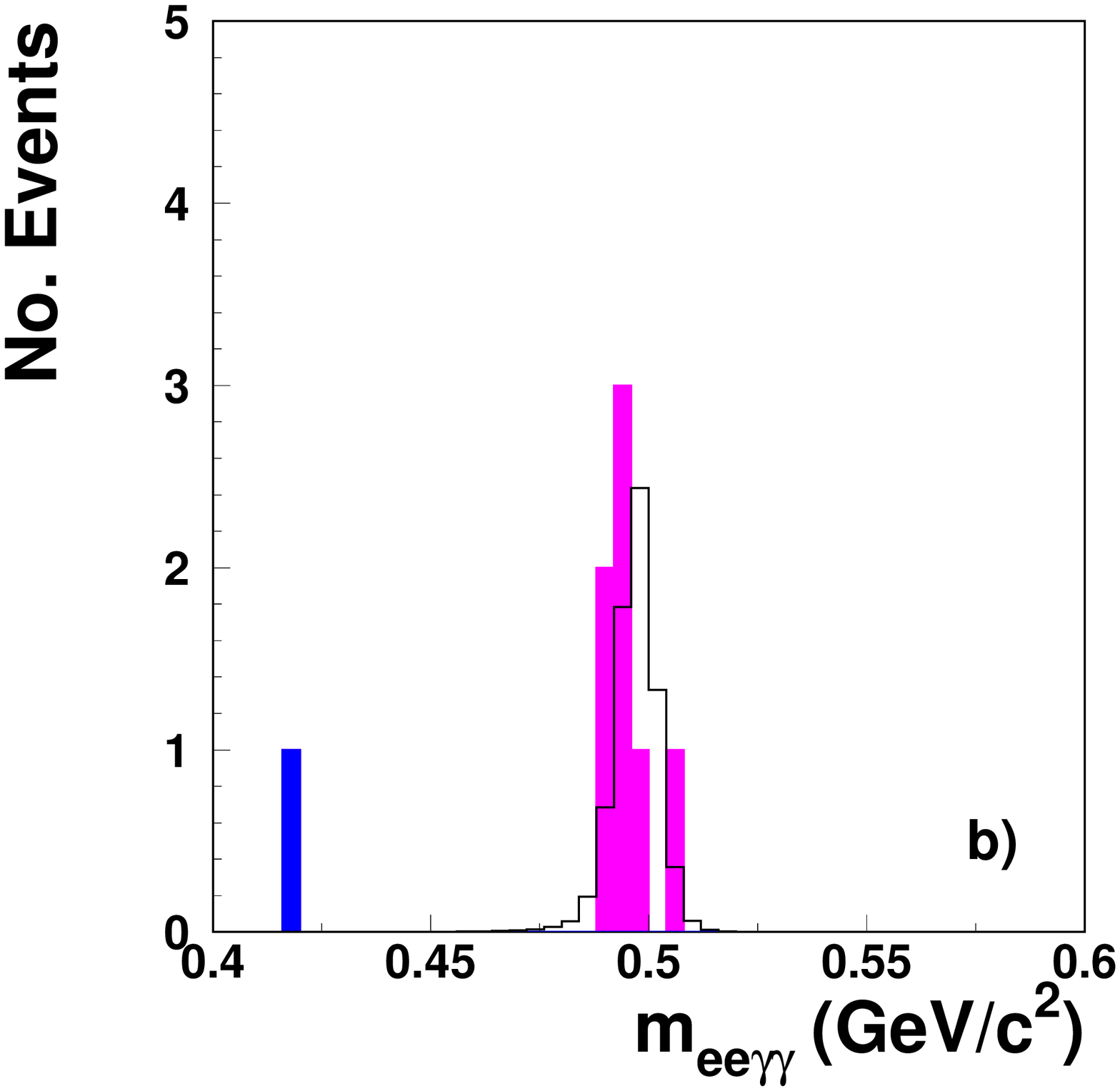}
    }
  }
%  \vspace{9pt}
%  \hbox{\hspace{1.35in} (a) \hspace{2.10in} (b)} 
%  \vspace{9pt}
\caption{$m_{\gamma \gamma}$ (a)
and $\meegg$ distributions (b) for the 7 events found
in the signal region. The expected 
Gaussian mass resolutions are superimposed (solid line).}
\label{cand}
\end{figure}

 Fig.~\ref{cand} shows the $m_{\gamma \gamma}$
and the $m_{e^+ e^- \gamma \gamma}$  
distributions of the events 
compared to the detector mass resolutions.
In Table \ref{events}, some of the 
kinematical quantities for each event are summarised.

In order to calculate the $\kspiee$ acceptance,
the amplitude for the decay was needed. This was 
taken from the Chiral Perturbation 
Theory prediction given in~\cite{bib:ambrosio}, and is of the form:

\begin{equation}
A[K(k) \rightarrow \pi(p) e^+ (p_+) e^- (p_-)] =  {\frac {-e^2}
{{m_{\text {\em K}}}^2 (4\pi)^2}} W(z) (k+p)^{\mu} 
{\bar u}_l (p_-) \gamma_{\mu} v_l (p_+)
\label{eq:matrix}
\end{equation}

\par\noindent
where $k$, $p$, $p_+$ and $p_-$ are
the four-momenta of the kaon, pion, positron and electron 
respectively; $m_{\text {\em K}}$ is the kaon mass; 
$W(z)$ is the electro-magnetic 
transition form factor, with
$z=(k-p)^2 / {m_{\text {\em K}}}^2$.
As a consequence of gauge invariance, the form factor 
dependence on $z$
vanishes to lowest order
%in the low-energy expansion in $z$
and therefore can be represented as a polynomial. 
For $K_S$ decays,
the form factor $W(z)$ was approximated to 
$W(z) \sim  a + b \times z $ \cite{bib:ambrosio}.

The $a$ and $b$ parameters have recently been measured for charged kaons,
and the ratio $a/b$ found to be 1.12~~\cite{bib:prl83}.

\begin{figure}[hbtp]
  \vspace{9pt}
  \centerline{\hbox{ \hspace{0.0in} 
    \epsfxsize=3.0in
    \epsffile{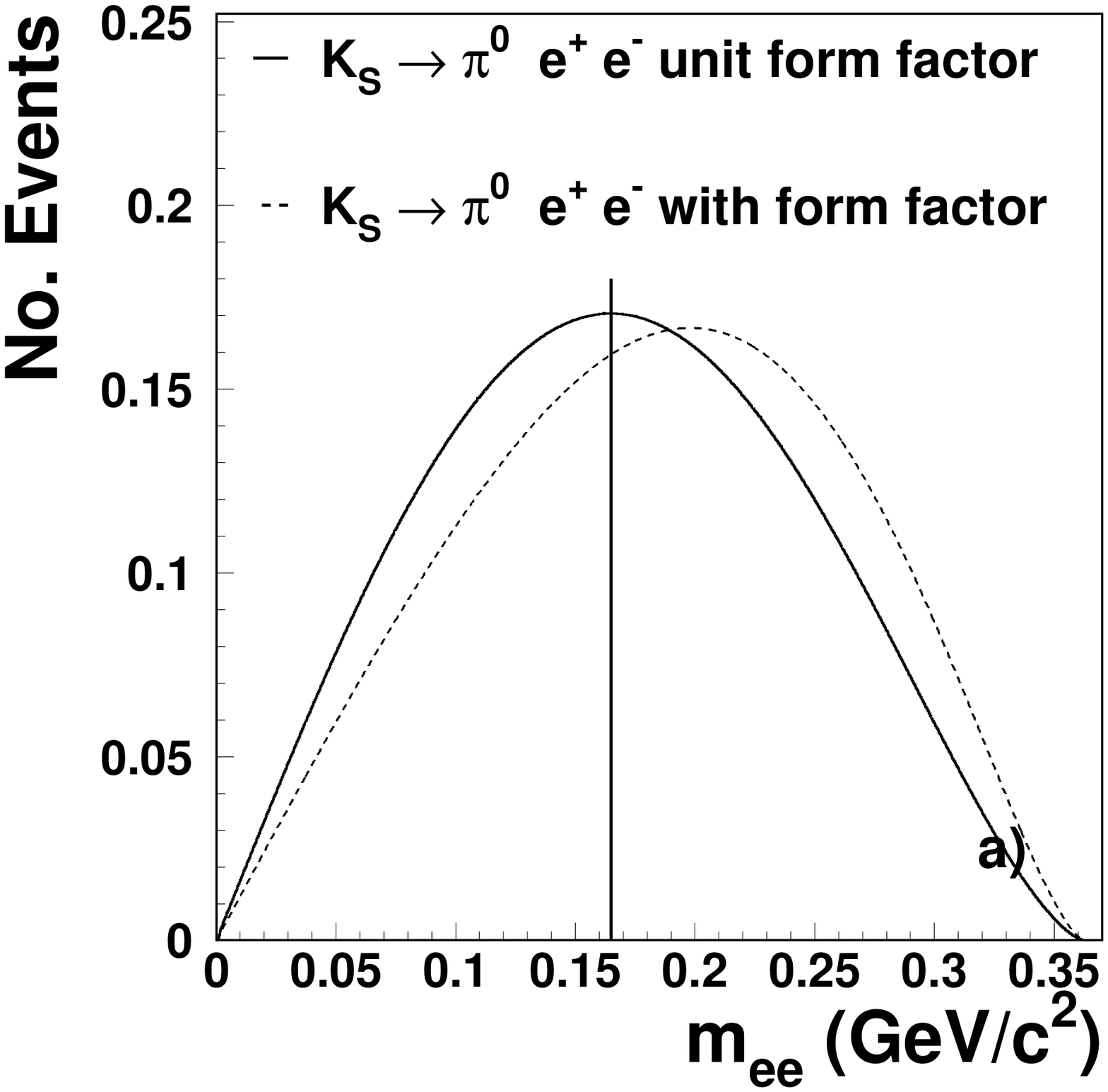}
    \hspace{0.25in}
    \epsfxsize=3.0in
    \epsffile{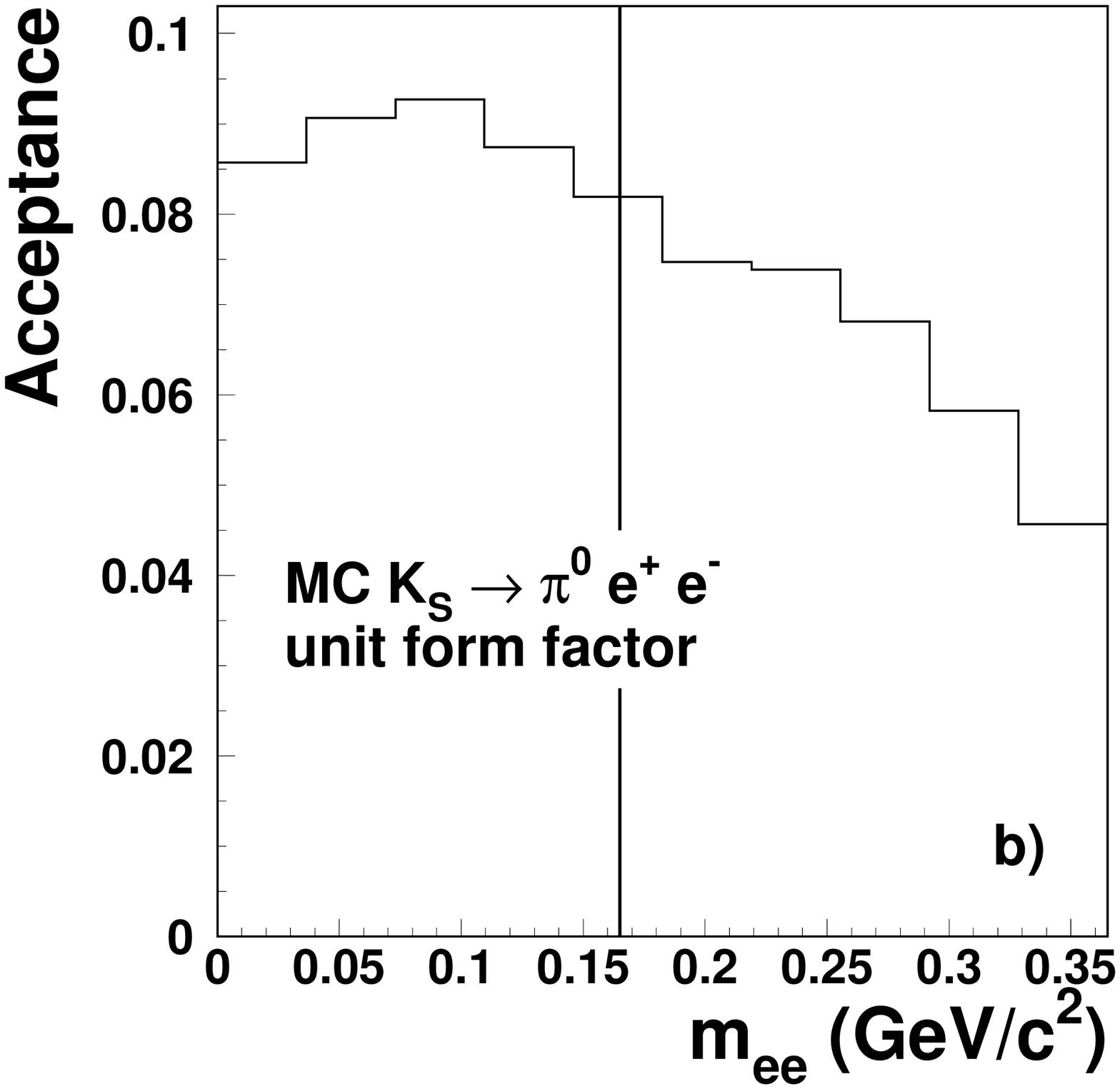}
    }
  }
%  \vspace{9pt}
%  \hbox{\hspace{1.35in} (a) \hspace{2.10in} (b)} 
%  \vspace{9pt}
\caption{$m_{ee}$ distributions from Monte Carlo, with
and without the form factor (a); acceptance as a function
of $m_{ee}$ (b).
}
\label{accep}
\end{figure}

The $m_{ee}$ distributions resulting from $W(z)=1$
and $W(z)= a + b \times z $ are shown
in fig. \ref{accep}.a.

The overall \kspiee acceptance depends on the form factor.
To remove this form factor dependence,
an acceptance was calculated for each event 
using fig. \ref{accep}.b, where the acceptance is given as
a function of $m_{ee}$, using $W(z)=1$.
The values are given in
Table \ref{events}.
The average geometrical acceptance of the 7 events is 0.15,
while the average analysis efficiency is 0.44,  which results
in an average efficiency of $0.066\pm 0.004$.
%The acceptance as a function ,
%is shown in  .

From the \kspipid\ flux and the signal of 6.85 events,
the branching ratio for $m_{ee} > 0.165$ GeV/\it{c}$^2$ \rm was computed:

$BR(\kspiee, \ m_{ee} > 0.165$ GeV/{\it c}$^2) = 
(3.0^{+1.5}_{-1.2} (\text{stat}) 
\pm 0.2 (\text{syst}) ) \times 10^{-9}.$

The quoted uncertainties correspond to a 68.27\% confidence 
level~\cite{bib:feldman}.
The systematic uncertainty includes the uncertainty of the
flux measurement and of the acceptance.
% variations.

%To extrapolate the branching ratio to the full
%$m_{ee}$ spectrum, the acceptance
%calculated with unit form factor has been used, obtaining:

%$$ BR(\kspiee) = (5.8^{+2.8}_{-2.3} (\text{stat}) 
%\pm 0.45 (\text{syst})) \times 10^{-9}.$$

\section{Discussion}

In Chiral Perturbation theory the $BR(\kspiee)$ is related
to the parameter $a_S$, which measures the strength 
of the indirect CP-violating term in \klpiee decay
as explained in \cite{bib:ambrosio} and eq.~\ref{eq:as}.

Using a vector matrix element with no form factor dependence,  
the measured branching ratio was extrapolated to the full
$m_{ee}$ spectrum to obtain:  

\begin{equation*}
BR(\kspiee) = (5.8^{+2.8}_{-2.3} (\text{stat})
\pm 0.8 (\text{syst})) \times 10^{-9}
\end{equation*}

The systematic error is dominated by the uncertainty
in the extrapolation due to the form factor dependence.
 
It was then possible to extract the parameter
$|a_S|$:

\begin{equation*}
|a_S| = (1.06 ^{+0.26} _{-0.21} (\text{stat}) \pm 0.07 (\text{syst}))
\end{equation*}

\begin{figure}[hbtp]
  \vspace{9pt}
  \centerline{\hbox{ \hspace{0.0in} 
    \epsfxsize=3.0in
    \epsffile{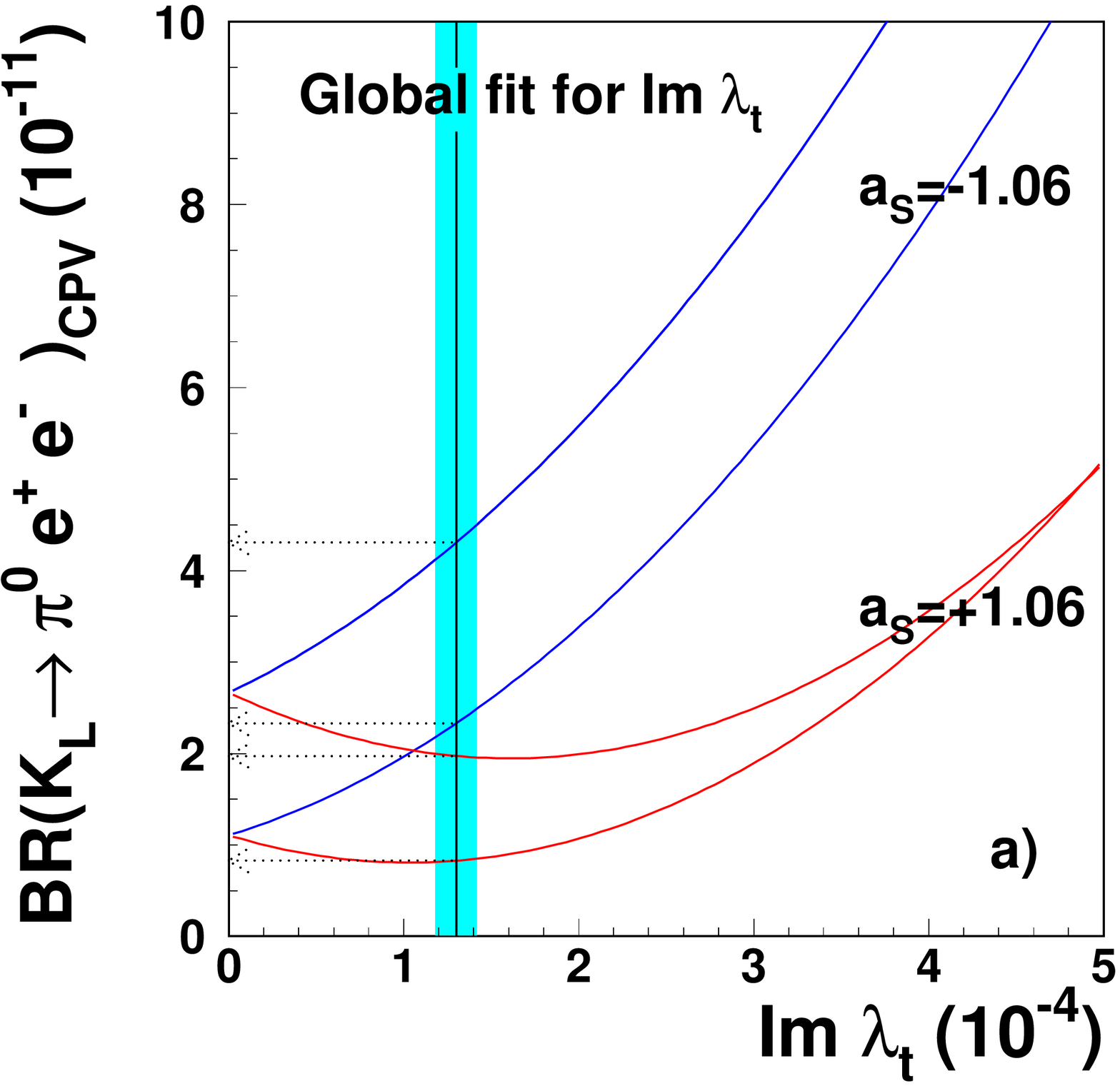}
    \hspace{0.25in}
    \epsfxsize=3.0in
    \epsffile{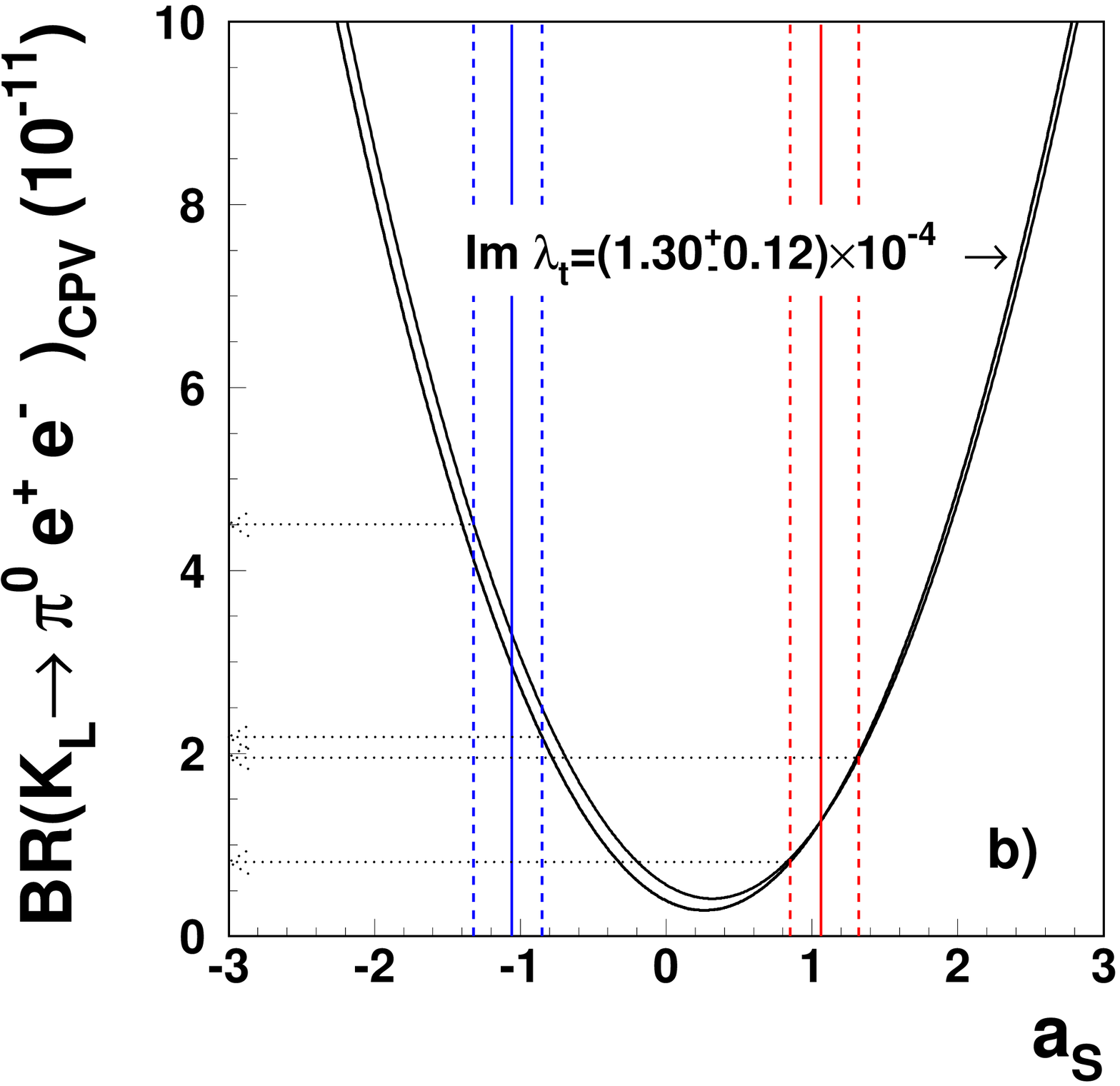}
    }
  }
%  \vspace{9pt}
%  \hbox{\hspace{1.35in} (a) \hspace{2.10in} (b)} 
%  \vspace{9pt}
\caption{Branching fraction of \klpiee as a function of $ Im(\lambda_t)$ (a), 
and as a function of the parameter $a_S$ (b).}
\label{implication}
\end{figure}

The measurement of $a_S$ allows the branching ratio 
{\rm BR}(\klpiee) to be predicted as
a function of $Im(\lambda_t)$ to within a sign ambiguity (see 
eq.~\ref{eq:interference}). The effect of the sign ambiguity 
can be seen in fig.~\ref{implication}.a.
%This is shown in fig.~\ref{implication}.a.

Alternatively, as shown in fig.~\ref{implication}.b,
by using the global fit value for
$Im(\lambda_t)=(1.30\pm 0.12)\times 10^{-4}$ obtained
from b-decay ~\cite{bib:lambdat}, 
{\rm BR}(\klpiee) can be expressed as function of $|a_S|$.

Using the measured value of $|a_S|$ and the global fit for
$Im(\lambda_t)$, eq.~\ref{eq:interference} reduces to:

\begin{displaymath}
{\rm BR}(\klpiee)_{\text{CPV}} \simeq 
(17.2_{\text{indirect}} \pm 9.4_{\text{interference}} + 4.7_{\text{direct}})
\times 10^{-12} .
\end{displaymath}

The CP conserving component 
can be obtained from the study of the $\klpigg$ decay.
A measurement made by the KTeV collaboration ~\cite{bib:ktev} found
$BR(\klpiee)_{CPC} = (1 - 2) \times 10^{-12}$. A more recent
measurement quoted
$BR(\klpiee)_{CPC} = 0.47^{+0.22}_{-0.18}\times 10^{-12} $~\cite{bib:piogg}
suggesting that the CP-conserving component is negligible.
 
%is therefore negligible and the total
%branching fraction can be estimated as 
%${\rm B}(\klpiee)=(1-4)\times 10^{-11}$.

Given the measured value of $a_S$ the direct CP violated component
predicted from the Standard Model
is small with respect to the indirect component.
If the sign of $a_S$ turns out to be negative then
BR($\klpiee$) retains some sensitivity to ${\rm Im}(\lambda_t)$ through
the interference term.

\section*{Acknowledgements}
It is a pleasure to thank the technical staff of 
the participating laboratories,
universities and affiliated computing centres for their efforts in the 
construction of the NA48 apparatus, in the operation of the experiment, and in 
the processing of the data.
We are also grateful to Giancarlo D'Ambrosio, Gino Isidori, and
Antonio Pich for 
useful discussions.

% \end{linenumbers}

%%%%%%%%%%%%%%%%%%%%%%%%%%%%%%%%%%%%%%%%%%%%%%%%%%%%%%%%%%

%\end{linenumbers}
\newpage

%%%%%%%%%%%%%%%%%%%%%%%%%%%%%%%%%%%%%%%%%%%%%%%%%%%%%%%%%%

%\begin{table}
%%\begin{center}
%\hspace*{-2cm}
%\begin{tabular}{|l|c|c|}
%\hline
%Source     &  control region  &  signal region \\
%\hline
%$K_S$  &    &   \\
%\hline
%$\Xi^0$    &    &    \\
%\hline
%%$K_L \rightarrow e^+ e^- \gamma \gamma$  &    &   \\
%\hline
%\end{tabular}
%\hspace*{0.5cm}
%\caption{Principle background sources.}
%\label{backphy}
%\end{center}
%\end{table}

\begin{table}
\begin{center}
\hspace*{-2cm}
\begin{tabular}{|l|c|c|}
\hline
Source     &  control region  &  signal region \\
\hline
\kspidpid  &  0.03  & $< 0.01$  \\
\hline
\kleegg    &  0.11  & 0.08   \\
%Phys bkgrd tot  &  0.14   &  0.08 \\
\hline
Accidentals & 0.19  & 0.07 \\
\hline\hline
Total background & 0.33  & 0.15 \\
\hline
\end{tabular}
\hspace*{0.5cm}
\caption{ Summary of the background estimate.}
\label{back}
\end{center}
\end{table}

\begin{table}
\begin{center}
\hspace*{-2cm}
\begin{tabular}{|c|c|c|c|c|}
\hline
 Event no. &  $K_S$ momentum & $\tau / \tau_{S}$ & $m_{ee}$ & Acceptance \\
\hline 
 & GeV/{\it c} &  & GeV/{\it c}$^2$ & \\
\hline
1 & 84.6  & 0.74 & 0.291  & 0.058 \\
2 & 128.2 & 0.50 & 0.267  & 0.066 \\
3 & 114.1 & 1.02 & 0.173  & 0.084 \\
4 & 83.9  & 2.09 & 0.272  & 0.066 \\ 
5 & 130.8 & 1.46 & 0.303  & 0.052 \\
6 & 121.2 & 1.49 & 0.298  & 0.058 \\
7 & 94.2  & 1.64 & 0.253  & 0.075 \\
\hline
\end{tabular}
\hspace*{0.5cm}
\caption{ Kinematical quantities of the seven events
found in the signal region.}
\label{events}
\end{center}
\end{table}

\newpage
%%%%%%%%%%%%%%%%%%%%%%%%%%%%%%%%%%%%%%%%%%%%%%%%%%%%%%%%%%%%

%\begin{figure}[hbt]
%\begin{center}
%\mbox{\epsfig{file=plot/ksklbeams_2001.eps,width=12cm}}
%\caption{View of the NA48 beam line}
%\label{beam}
%\end{center}
%\end{figure}

%\begin{figure}[hbt]
%\vspace{1.0cm}
%\begin{center}
%\mbox{\epsfig{file=plot/na48_det_2d_modified.eps,width=12cm}}
%\caption{Plan view of the NA48 detector}
%\label{detector}
%\end{center}
%\end{figure}

%\begin{figure}[hbt]
%\begin{center}
%\mbox{\epsfig{file=plot/mee_minus_3.44_partial.eps,width=8cm}}
%\caption{Distributions of $m_{ee}$ after all the cuts have been applied,
%for opposite sign minus same sign events.}
%\label{pi0ee}
%\end{center}
%\end{figure}

%\begin{figure}[hbtp]
%  \vspace{9pt}
%  \centerline{\hbox{ \hspace{0.0in} 
%    \epsfxsize=3.0in
%    \epsffile{plot/mkaon_ot.eps}
%    \hspace{0.25in}
%    \epsfxsize=3.0in
%    \epsffile{plot/mpion_ot.eps}
%    }
%  }
%  \vspace{9pt}
%  \hbox{\hspace{1.35in} (a) \hspace{2.10in} (b)} 
%  \vspace{9pt}
%\caption{Scatter plot of $\meegg$ (a) and $m_{\gamma \gamma}$ (b)
%as a function of the difference between the average time of the two
%clusters associated with tracks and the average time of the 
%two neutral clusters, $\Delta t$,
%for 2002 data. The region of event time less than 3 ns
%has been excluded. The regions of $3 \sigma$ and $6 \sigma$ are shown.}
%\label{aot}
%\end{figure}

\begin{thebibliography}{00}
\bibitem{sehgal} L.~M.~Sehgal, Nuclear Physics B19 (1970) 445.

\bibitem{bib:triangle} G.~Ecker, A.~Pich, and E.~de Rafael, 
Nuclear Physics B291 (1987) 692.

\bibitem{bib:prades} C.~Bruno and J.~Prades, 
Zeitschrift f\"{u}r Physik C57 (1993) 585. 

%\cite{Ecker:fm}
%\bibitem{Ecker:fm}
%G.~Ecker, A.~Pich and E.~de Rafael,
%``K0 $\to$ Pi0 Gamma Gamma Decays In Chiral Perturbation Theory,''
%Physics Letters B189 (1987) 363.
%%CITATION = PHLTA,B189,363;%%


%\cite{Ecker:1987hd}
\bibitem{Ecker:1987hd}
G.~Ecker, A.~Pich and E.~de Rafael,
%``Radiative Kaon Decays And CP Violation In Chiral Perturbation Theory,''
Nuclear Physics B303 (1988) 665.
%%CITATION = NUPHA,B303,665;%%

\bibitem{bib:gabbiani}
J.F.~Donoghue, F.~Gabbiani, Physical Review D51 (1995) 2187.

\bibitem{bib:ambrosio} G.~D'Ambrosio, G.~Ecker,
G.~Isidori, and J.~Portoles, Journal of High Energy Physics 08 (1998) 004.

%\cite{D'Ambrosio:1998yj}
%\bibitem{D'Ambrosio:1998yj}
%G.~D'Ambrosio, G.~Ecker, G.~Isidori and J.~Portoles,
%``The decays K $\to$ pi l+ l- beyond leading order in the chiral  expansion,''
%Journal of High Energy Physiscs 08 (1998) 004.
%[arXiv:hep-ph/9808289].
%%CITATION = HEP-PH 9808289;%%

%\cite{Alavi-Harati:2000sk}
\bibitem{Alavi-Harati:2000sk}
A.~Alavi-Harati et al.,
%``Search for the decay K(L) $\to$ pi0 e+ e-,''
Physical Review Letters  86 (2001) 397.
%[arXiv:hep-ex/0009030].
%%CITATION = HEP-EX 0009030;%%

%\cite{Lai:2001jf}
\bibitem{Lai:2001jf}
A.~Lai et al.,
%``Search for the decay K(S) $\to$ pi0 e+ e-,''
Physics Letters B514 (2001) 253.
%%CITATION = PHLTA,B514,253;%%

%\cite{Isidori:2001nd}
\bibitem{Isidori:2001nd}
G.~Isidori,
%``Rare decays: Theory vs. experiments,''
International Journal of  Modern Physics A17 (2002) 3078.
%[arXiv:hep-ph/0110255].
%%CITATION = HEP-PH 0110255;%%


\bibitem{bib:epsi} J.R.~Batley et al., Physics Letters B544 (2002) 97. 

\bibitem{bib:unal} G.~ Unal, NA48 Collaboration, in: IX International 
Conference on Calorimetry, October 2000, Annecy, France, hep-ex/0012011.

\bibitem{bib:trigneut} G.~Barr et al., Nuclear Instruments and Methods in
Physics Research A485 (2002) 676.

\bibitem{bib:geant} GEANT Detector Description and Simulation
Tool, CERN Program Library Long Write-up W5013 (1994).

%\bibitem{bib:PDG} D.E.~Groom et al., European Physical Journal C15 (2000) 1. 
\bibitem{bib:PDG} K.~Hagiwara et al., Physical Review D 66 (2000) 1. 

\bibitem{bib:prl83} R.~Appel et al., Physical Review Letters 83 (1999) 4482.

\bibitem{bib:feldman} G.J.~Feldman and R.D.~Cousins, 
Physics Review D57 (1998) 3873.

\bibitem{bib:lambdat} S.~H.~Kettell, L.~G.~Landsberg,
and H.~Nguyen, hep-ph/0212321.

\bibitem{bib:ktev} A.~Alavi-Harati et al., Physical Review Letters 83 (1999) 917.

\bibitem{bib:piogg} A.~Lai, et al., Physics Letters B536 (2002) 229.


% notes:

% \bibitem{label2}
% If there is a note, it should come last:
% \bibitem{label3} \note
% \end{subbibitems}


\end{thebibliography}
\end{document}